\documentclass[journal,12pt,onecolumn,draftclsnofoot]{IEEEtran}
\usepackage{times}
\usepackage[final]{graphicx}
\usepackage{textcomp}
\usepackage[reqno]{amsmath}
\usepackage{amsfonts}
\usepackage{times,amsmath,epsfig}
\usepackage{latexsym,amssymb}
\usepackage{cite}
\usepackage{psfrag}
\usepackage{subfig}
\usepackage{stfloats}
\usepackage{amsmath}
\usepackage{amssymb}
\usepackage{array}
\usepackage[noend]{algorithmic}
\usepackage{algorithm}
\usepackage{bm}
\usepackage{color}
\usepackage{multicol}
\usepackage{multirow}
\usepackage{setspace}
\usepackage{tabularx}
\usepackage[T1]{fontenc}
\newtheorem{rem}{Remark}
\newtheorem{lem}{Lemma}
\newtheorem{prob}{Problem}
\newcommand{\PreserveBackslash}[1]{\let \temp =\\#1 \let \\ = \temp}
\newcolumntype{C}[1]{>{\PreserveBackslash\centering}p{#1}}
\newcolumntype{R}[1]{>{\PreserveBackslash\raggedleft}p{#1}}
\newcolumntype{L}[1]{>{\PreserveBackslash\raggedright}p{#1}}

\newlength{\figwidth}
\begin{document}
\baselineskip 19.5 pt

\title{ Energy-efficient Resource Allocation for Mobile Edge Computing Aided by Multiple Relays}
\author{
Xiang Li, Rongfei Fan, Han Hu, Ning Zhang, and Xianfu Chen
\thanks
{
%
X. Li, R. Fan, and H. Hu are with the School of Information and Electronics, Beijing Institute of Technology, Beijing 100081, P. R. China. (\{lawrence,fanrongfei,hhu\}@bit.edu.cn,).
N. Zhang is with the Department of Electrical and Computer Engineering, University of Windsor, Windsor, ON, N9B 3P4, Canada. (ning.zhang@uwindsor.ca).
X. Chen is with the VTT Technical Research Centre of Finland, Oulu 90571, Finland (xianfu.chen@vtt.fi).

}
}

\maketitle

\begin{abstract}
In this paper, we study a mobile edge computing (MEC) system with the mobile device aided by multiple relay nodes for offloading data to an edge server. Specifically, the modes of decode-and-forward (DF) with time-division-multiple-access (TDMA) and frequency-division-multiple-access (FDMA), and the mode of amplify-and-forward (AF) are investigated, which are denoted as DF-TDMA, DF-FDMA, and AF, respectively. Our target is to minimize the total energy consumption of the mobile device and multiple relay nodes through optimizing the allocation of computation and communication resources. Optimization problems under the three considered modes are formulated and shown to be non-convex. For DF-TDMA mode, we transform the original non-convex problem to be a convex one and further develop a low computation complexity yet optimal solution. In DF-FDMA mode, with some transformation on the original problem, we prove the mathematical equivalence between the transformed problem in DF-FDMA mode and the problem under DF-TDMA mode. In AF mode, the associated optimization problem is decomposed into two levels, in which monotonic optimization is utilized in upper level and successive convex approximation (SCA) is adopted to find the convergent solution in the lower level. Numerical results prove the effectiveness of our proposed methods under various working modes.
\end{abstract}


\section{Introduction} \label{s:intro}
Recent years have witnessed a surging demand for computation in a wide range of emerging mobile applications, such as image recognition, virtual reality (VR), and augmented reality (AR) \cite{Shi_survey}, which are computationally intensive in general.
These applications pose great challenges for mobile devices in terms of high energy consumption if the computation task is computed at local or long latency when the computation task is offloaded to a remote cloud center.
Mobile edge computing (MEC) emerges as a promising solution to tackle the above problem \cite{architecture_survey}.
In a MEC system, an edge server rich in computation capacity is implemented in the vicinity of mobile devices, usually on the base station (BS).
Hence a mobile device can offload the data for computing at the BS with a lower latency. The BS can complete computing the offloaded data shortly and save the mobile device from consuming too much energy on local computing.

In terms of offloading the data for computing from a mobile device to an BS, partial offloading is the most popular way \cite{Taleb_survey}, in which the data for computing can be partitioned into multiple parts and processed at multiple sites in parallel.
With partial offloading in a MEC system, the computation of the task is composed of two parts: the computation at local and the computation at the BS, both of which will involve energy consumption at the mobile device.
The two parts will be hereinafter referred to as local computing and edge computing, respectively.
Specifically, the energy of mobile device will be consumed partly on mobile device's CPU when local computing is executed, and will be consumed partly on wireless transmission when the data for edge computing is offloaded \cite{Huang_survey}.
The energy consumption of the mobile device and the delay for completing the computation task are two important performance metrics \cite{ref_97}.
The general design goals of a MEC system include minimizing the energy consumption of the mobile device or BS \cite{vm}, the time delay for task completion \cite{Wu_noma}, or weighted sum of above two performance metrics \cite{fd}.
To achieve the research goals, the allocation of both communication and computation resource are performed in literature.
The communication resource primarily involves the amount of data for offloading \cite{ref_112}, mobile device's transmit power \cite{Xu_d2d}, bandwidth or time slot duration in data offloading \cite{ref_84}.
The computation resource mainly includes the CPU frequency of the mobile device and the BS \cite{ref_133}.

The research on the allocation of communication resource and computation resource is firstly investigated for a MEC system with single mobile device.
As an example of early research in this field, reference \cite{ref_97} studied the minimization of energy consumption when one mobile device partially offloads its task to the BS. The proportion of offloaded data, frequency of local CPU for local computing and transmit power for task offloading are jointly optimized.
The research on the allocation of communication resource and computation resource is then extended to the MEC system with multiple mobile devices.
To guarantee successful transmission, different communication protocols for multi-user access to offload the task is considered by various research.
In \cite{ref_84}, multiple mobile devices access into the BS via the way of either time-division multiple access (TDMA) or orthogonal frequency-division multiple access (OFDMA) with the purpose of minimizing the energy consumption of all the mobile devices. In TDMA scenario, the slot duration for data offloading along with the amount of data for offloading are optimized for every mobile device, subject to the constraints of BS's computation capability and task completion latency. A convex optimization problem is formulated and simple solution can be generated by inspecting a derived offloading priority function.
In OFDMA scenario, channel allocation is executed together with the allocation of offloaded data for every mobile device.
A mixed-integer optimization problem is formulated and sub-optimal low-complexity solution is presented.
Unlike \cite{ref_84}, \cite{Cui_wpt} considers a wirelessly powered MEC system supporting multiple mobile devices. In such a system, the BS wirelessly transmits power, which is referred to as wireless power transfer (WPT) technology, to multiple mobile devices through multi-antenna, and every mobile device offloads their data back to the BS by exhausting harvested energy via TDMA.
To minimize overall energy consumption of the BS, the energy transmit beamforming at the BS, the CPU frequency, the amount of data for offloading and duration of time slot for every mobile device are optimized.

In existing literature on the MEC system with single or multiple mobile devices, the cooperation among multiple mobile devices has been seldom considered. In scenarios where the channel is not ideal due to long distance or deep fading, it would be helpful if other idle mobile devices in the vicinity of both the mobile device and the BS can help to relay the offloaded data.
\cite{ref_127} considers a single user MEC system with one relay node.
The relay node can help the mobile device to not only forward information but also compute the offloaded data.
With some data left for local computing, the mobile device divides the rest of data into two parts: one part is transmitted to the relay node for computing at the relay node, and the other part is offloaded to the BS with the aid of the relay node for computing at the BS. This procedure is completed through time-division manner. The duration of every time slot, the amount of data for each part of computing, and the transmit power on every communication link are optimized together in order to minimize the energy consumption of mobile device and relay node.
\cite{wpt_coop} considers a single-user WPT-driven MEC system with one relay node.
Different from \cite{ref_127}, the relay node does not help the mobile device to compute but has its own task to complete, which would be partly assisted by BS.
Partial offloading is assumed for both mobile device and relay node.
The mode of harvest-then-offload is utilized, in which both the mobile device and the relay node first harvest energy from the BS through WPT, then the mobile device offloads part of its data to the BS with the aid of relay node, and at last the relay node offloads part of its own data to the BS.
For such a system, the amount of data to offload for both the mobile device and relay node, the duration of every phase, and the transmit power over every wireless link are optimized for minimizing energy consumption of the BS.


In aforementioned related works, only single relay node is considered. In a real MEC system, e.g., 5G network implemented with MEC technology, there may exist multiple relay nodes between the mobile device and the BS.
Exploiting the diversity from these multiple relay nodes can achieve performance gain over the case with single relay node.
In this paper, for the first time we investigate a MEC system with multiple relay nodes between the mobile device and the BS.
With the purpose of minimizing the energy consumption of both the mobile device and the relay nodes while satisfying the time delay requirements for completing the computation task,  two general working modes in a relay network are investigated: Decode-and-forward (DF) mode and amplify-and-forward (AF) mode. In DF mode, the multiple relay nodes can access the mobile device and the BS in the way of time-division-multiple-access (TDMA) and frequency-division-multiple-access (FDMA), which are named as DF-TDMA mode and DF-FDMA mode, respectively.
For these three working modes, {we optimize the computation and communication resource allocation at the mobile device, multiple relays, and the BS, while satisfying the mobile device’s computation latency constraint.} The contributions are summarized as follows.
\begin{itemize}
	\item DF-TDMA: In this mode, the amount of data to offload, transmit power over every communication link, and the slot duration allocated to every communication link are optimized jointly. An optimization problem is formulated, which is non-convex. To get the global optimal solution, we transform it to be a convex optimization problem equivalently. To further reduce the computation complexity, the transformed convex optimization problem is decomposed into two levels. The amount of data to offload is optimized in the upper level, while all the other variable are optimized in the lower level. With a series of analysis and transformation, the lower level optimization problem are transformed to be a linear programming problem and the upper level optimization problem can transformed to be an one-dimensional convex optimization problem, whose optimal solution can be found using Golden search method easily. Numerical result shows that our proposed method can achieve lower computation complexity compared with traditional interior point method, which is broadly used for solving a general convex optimization problem.
	
	\item DF-FDMA: In this mode, the amount of data to offload, transmit power over every communication link, and the  spectrum allocated to every communication link are optimized jointly. The associated optimization problem is still non-convex. By exploiting special properties beneath the formulated optimization problem, we find the hidden equivalence between the optimization problem in DF-FDMA mode and the transformed convex optimization problem in DF-TDMA mode.
Hence the optimization problem in DF-FDMA mode can be solved optimally and easily by following the same method for DF-TDMA mode.
	
	\item AF: In this case, the amount of data to offload, transmit power of the mobile device, and the amplifying coefficient on every relay are optimized jointly. An optimization is formulated, which is non-convex either. To find the solution, we decompose it into two levels, in the upper level of which the amount of data to offload is optimized and in the lower level of which all the rest variables are optimized. With steps of transformations and by utilizing successive convex approximation (SCA), convergent solution for the level optimization problem can be found. By resorting to monotonic optimization and polyblock algorithm, the optimal solution for the transformed upper level optimization problem can be found.
\end{itemize}

The rest of the paper is organized as follows.
System model is presented in Section \ref{s:system_model}.
Optimization problems for the three investigated working modes are formulated in Section \ref{s:problem_formulation}.
The solution for the formulated optimization problems associated with the three working modes are demonstrated in Section \ref{s:prob_solution}.
Numerical results are given in Section \ref{s:numerical_results}, followed by conclusion remarks in Section \ref{s:conclusion}.

\section{System Model} \label{s:system_model}

\begin{figure}
\centering
\subfloat [System scenario]{
\label{f:all}
\includegraphics[width=1.0 \figwidth]{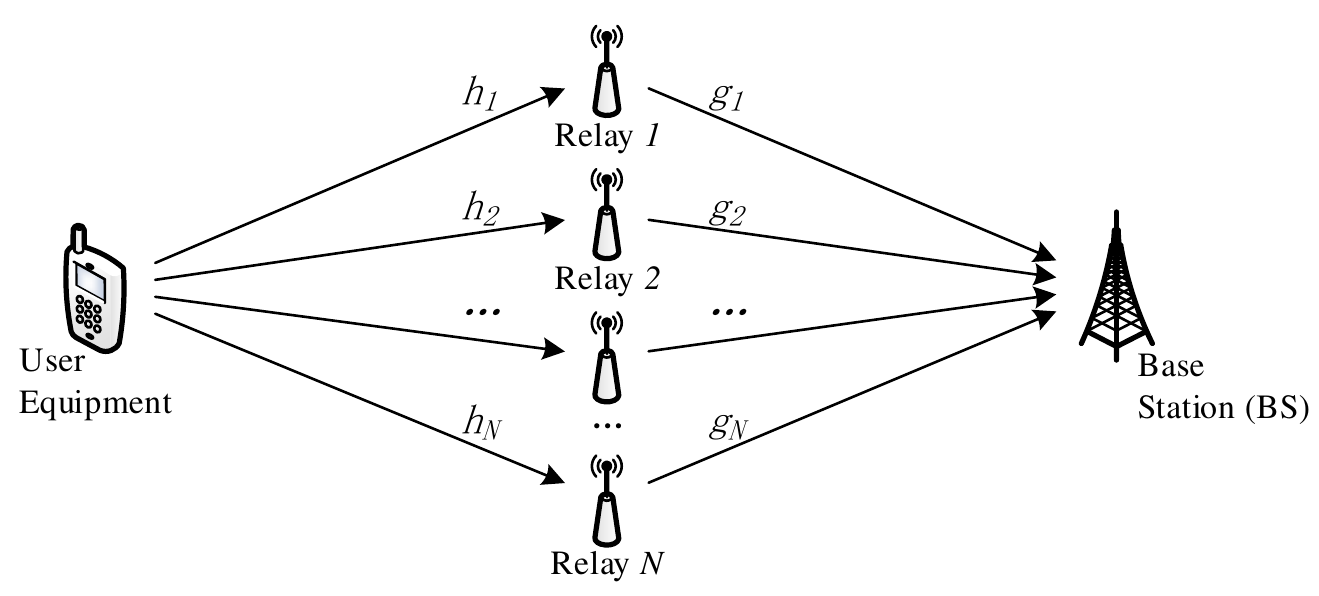}}
\subfloat [Frame structure]{
\label{f:slots}
\includegraphics[width=1.0 \figwidth]{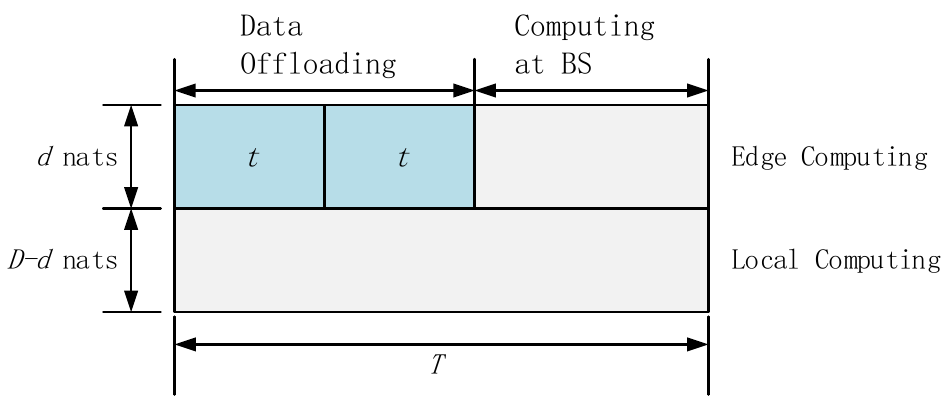}
}

\caption{System model.}
\end{figure}

Consider a MEC system with one mobile device, one edge server at a BS, and $N$ relay nodes, which are denoted as $n$th relay or relay $n$ for $n\in \mathcal{N}$ where $\mathcal{N} \triangleq \{1, 2, ..., N\}$, as shown in Fig.\ref{f:all}\footnote{The discussion of this paper can be easily extended to the case with multiple mobile devices by allocating every mobile device with an exclusive part of spectrum. In addition, the energy consumption of the relay nodes is not ignorable in case of multiple mobile devices, which has already been taken into account in this paper.}.
The mobile device has one computation task, say task $\mathcal{T}$, to complete.
The task $\mathcal{T}$, can be described by a three-tuple $\left(T, D, L\right)$.
In this tuple, $T$ indicates the maximal delay the mobile device can tolerate.
$D$ is the total amount of input data to process in order to complete task $\mathcal{T}$, which is in unit of nat for simplicity of presentation.
$L$ is the number of CPU cycles required for computing unit data nat.
Hence $LD$ represents the total amount of CPU cycles to process for completing task $\mathcal{T}$.

The task $\mathcal{T}$ is assumed to be separable, which means that it can be processed at more than one sites simultaneously.
To complete task $\mathcal{T}$ in an energy-efficient way, the mobile device can offload part of the data for computing to the BS, while the rest of data is left for local computing.
Suppose the amount of data to offload to the BS is $d \in [0, D]$. Then the amount of data for local computing is $\left(D-d\right)$.
For the data offloaded to the BS, after receiving the offloaded data, the BS will first process it and then return computational results to the mobile device. Since the computational results are always of small data size and the BS (who is generally rich in power supply) can easily achieve high data rate, the time for feeding back the computational results to the mobile device is ignorable and omitted \cite{rate_max_wpt}.

Due to blockage or deep fading, there is no direct link between the mobile device and the BS.
Hence these $N$ relay nodes help the mobile device to offload data to the BS.
Denote the channel gain between the mobile device and $n$th relay as $h_n$, and the channel gain between $n$th relay and BS as $g_n$ for $n \in \mathcal{N}$.
Both $h_n$ and $g_n$ for $n\in \mathcal{N}$ are block faded, which means that these channel gains are stable within the duration of one fading block and vary randomly and independently over different fading blocks.
Without loss of generality, the length of one fading block, which is generally at the scale of coherence time, is larger than $T$. This assumption is reasonable in slow-fading environment.
Note that the values of $h_n$ and $g_n$ for $n\in \mathcal{N}$ can be measured at the beginning of one fading block with negligible time overhead.
Suppose the system bandwidth is $W$, which is not larger than coherence bandwidth. Hence both $h_n$ and $g_n$ for $n\in \mathcal{N}$ are flat within the system bandwidth $W$.

As shown in Fig. \ref{f:slots}, task is divided and executed at local and at the BS simultaneously within time $T$. At local, $\left(D-d\right)$ nats of data is computed and finished before deadline. While at the BS, a time of $2t$ is preserved for data offloading.
To finish the transmission, these $N$ relay nodes work in a half-duplex manner. Specifically, time for data offloading is divided into two equal-length phases. Hence each phase occupies a time of $t$.
In the first phase, the mobile device transmits the offloaded data to the relays, whereas in the second phase, every relay forwards the received signal or decoded information to the BS.
Three working modes for the relay nodes are investigated, which are given as follows.

\begin{itemize}
\item DF-TDMA: 
In this mode, these $N$ relay nodes utilize DF and work in the way of TDMA.
This indicates that every relay node first decodes the received signal from the mobile device and then forwards the decoded information to the BS, which will happen in the first phase and second phase, respectively. In addition, these relay nodes are orthogonal over time.
As shown in Fig. \ref{f:tdma}, we suppose that these $N$ relays work on the common bandwidth $W$, and $n$th relay occupies a time window of $t_n$ both in the first phase and the second phase for $n\in \mathcal{N}$.
Then there are
\begin{equation} \label{e:tdma_t_positive}
t_n \geq 0, \forall n \in \mathcal{N},
\end{equation}
and
\begin{equation} \label{e:tdma_sum_t_n}
\sum_{n=1}^N t_n \leq t.
\end{equation}
For $n\in \mathcal{N}$, define the transmit power from mobile device to $n$th relay as $P_n$ and the transmit power from $n$th relay to BS as $Q_n$, respectively, both of which are larger than zero.
Denote the power spectral density (PSD) of background noise as $\sigma^2$, then
the amount of data transmitted through $n$th relay, defined as $D_n^T$, is the minimum value between the first phase and the second phase, i.e.,
$D_n^{T}=\min \Bigg(t_n W\ln \left(1+\frac{P_nh_n}{\sigma^2W} \right),
t_n W\ln \left(1+\frac{Q_ng_n}{\sigma^2W}\right)\Bigg), \forall n \in \mathcal{N}.$
Thus, to offload $d$ nats, there is
\begin{equation} \label{e:tdma_sum_data}
d \leq \sum \limits_{n=1}^N D_n^T.
\end{equation}


\begin{figure}
\centering
\subfloat [TDMA]{
\label{f:tdma}
\includegraphics[width=0.99 \figwidth]{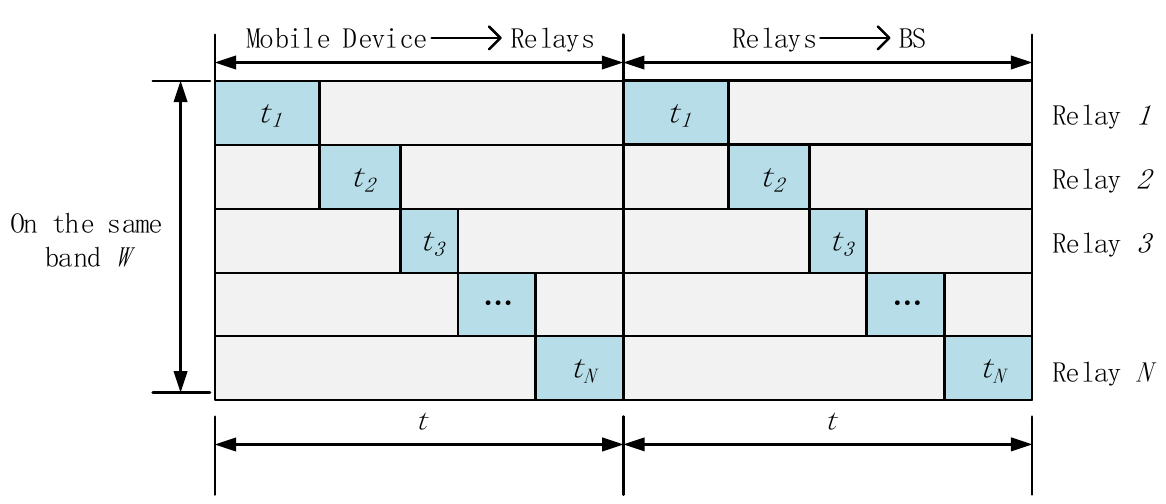}}
\subfloat [FDMA]{
\label{f:fdma}
\includegraphics[width=0.90\figwidth]{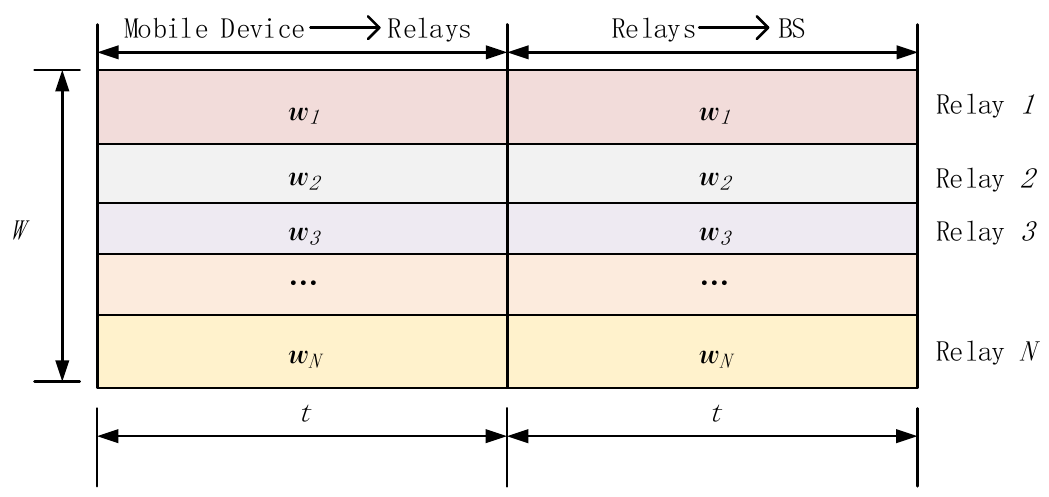}
}
\caption{Communication resource allocation.}
\end{figure}

\item DF-FDMA: 
In this mode, these $N$ relay nodes also utilize DF but work in the way of FDMA.
As shown in Fig. \ref{f:fdma}, $n$th relay spans a time of $t$ and occupies the spectrum with the bandwidth $w_n$ both in the first phase and the second phase for $n\in \mathcal{N}$. Then there are
\begin{equation} \label{e:fdma_w_positive}
w_n \geq 0, \forall n \in \mathcal{N},
\end{equation}
and
\begin{equation} \label{e:fdma_sum_w_n}
\sum_{n=1}^N w_n \leq W.
\end{equation}
With the same definitions of $P_n$, $Q_n$ and $\sigma^2$ as in DF-TDMA mode, the amount of data transmitted through $n$th relay is
$D_n^{F}=\min \Bigg(tw_n\ln\left(1+\frac{P_nh_n}{\sigma^2w_n} \right),
tw_n\ln \left( 1+\frac{Q_ng_n}{\sigma^2w_n} \right)\Bigg), \forall n \in \mathcal{N}.$
Thus, to offload $d$ nats, there is
\begin{equation} \label{e:fdma_sum_data}
d \leq \sum \limits_{n=1}^N D_n^F.
\end{equation}


\item AF: 
In this mode, the mobile device will transmit the offloaded data to the relays in the first phase, whereas in the second phase, all the relays will amplify the received signal and then transmit it to the BS \cite{AF_duplex}.
In such a process, every relay will occupy all the available spectrum $W$ and time $t$ in the first phase or the second phase.
Suppose the transmitted signal of the mobile device with unit power is $s$, the power of transmitted signal is $P$, and the additive noise at $n$th relay is $z_n$ with variance being $\sigma^2W$, then the received signal at $n$th relay is
$m_n=\sqrt{h_nP} s + z_n, \forall n \in \mathcal{N}.$
Denote the amplifying coefficient of $n$th relay as $\beta_n$, where $\beta_n \geq 0$, and the additive noise at the BS as $z_0$ with variance being $\sigma^2W$, then the energy consumption of relay $n$ is
$\beta_n^2\left(Ph_n+\sigma^2W\right)t$,
and the received signal at the BS in the second phase, denoted as $r$, can be expressed as
\begin{equation} \label{e:af_bssignal}
r=\sum_{n=1}^N \sqrt{g_n}\beta_n\left(\sqrt{h_n P}s + z_n\right)+ z_0
=\sum_{n=1}^{N}\sqrt{h_ng_n P}\beta_n s +\sum_{n=1}^{N} \sqrt{g_n}\beta_n  z_n + z_0.
\end{equation}
According to the expression of $r$ in (\ref{e:af_bssignal}), the power of the signal from the mobile device can be written as
$P_s^A= P\left(\sum_{n=1}^{N} \sqrt{h_ng_n} \beta_n \right)^2,$
and the noise power can be given as
$P_{z}^{A}=\sigma^2W+\sigma^2W \sum_{n=1}^N g_n\beta_n^2.$
Hence the amount of data transmitted from the mobile device to the BS through all the relays in AF mode can be given as \cite{AF_capacity}
$D^{A} =tW\ln \left(1+\frac{P_{s}^{A}}{P_{z}^{A}}\right)
      =tW\ln \left(1+\frac{P\left(\sum_{n=1}^{N} \sqrt{h_ng_n} \beta_n \right)^2}{\sigma^2 W \left(1+\sum_{n=1}^N g_n\beta_n^2 \right)}\right).$
Then to offload $d$ nats, there is
\begin{equation} \label{e:AF_sum_data}
d \leq D^A.
\end{equation}

\end{itemize}

In terms of computation, two parts need to be considered, i.e., the computation at local and the computation at the BS. For the computation at local, the mobile device needs to complete computing of $L (D-d)$ CPU cycles within time $T$. According to \cite{ref_97},
the energy consumption of the mobile device in this case can be expressed to be
\begin{equation} \label{e:local_comp_energy}
E_{c}=\frac{\kappa L^3 \left(D-d\right)^3}{T^2},
\end{equation}
where $\kappa$ is a given coefficient depending on the physical structure of mobile device's CPU.

For the computation at the BS, the BS need to complete the computing $Ld$ CPU cycles within time $\left(T - 2t\right)$, which corresponds to the maximal left time for computing at the BS. Suppose the computation capacity allocated to the mobile device by the BS is $f_B$. Then there should be
\begin{equation} \label{e:edge_comp_freq}
\frac{Ld}{(T-2t)} \leq f_{B}.
\end{equation}

\section{Problem Formulation} \label{s:problem_formulation}
In this section, based on the introduced system model in Section \ref{s:system_model}, optimization problems under three relay modes, including DF-TDMA, DF-FDMA, and AF, are presented respectively.
For all these three modes, or say cases, our objective is to minimize the overall energy consumption of the mobile device and all the relays by adjusting the amount of data for offloading, time duration for offloading, transmit power of mobile device, and power consumption strategy at every relay node, while respecting the latency requirement for completing the task $\mathcal{T}$.

For the case of DF-TDMA, according to (\ref{e:local_comp_energy}),
the total energy consumption of the mobile device and $N$ relay nodes can be written as
$\sum_{n=1}^N \left(P_nt_n+Q_nt_n \right) +\frac{\kappa L^3 (D-d)^3}{T^2}.$
Then collecting the constraints in (\ref{e:tdma_t_positive}), (\ref{e:tdma_sum_t_n}), (\ref{e:tdma_sum_data}), (\ref{e:edge_comp_freq}), and imposing the non-negative restrictions on $P_n$, $Q_n$, $t_n$ for $n \in \mathcal{N}$ and box constraint on variable $d$, the optimization problem can be formulated as follows
\begin{prob} \label{p:tdma_first} \small
	\begin{subequations}
	\begin{align}
	\min_{\substack{d,\{t_n|n\in \mathcal{N}\}, \{P_n|n\in \mathcal{N}\}, \{Q_n|n\in \mathcal{N}\}}}
	&\sum_{n=1}^N \left(P_nt_n+Q_nt_n \right) +\frac{\kappa L^3 (D-d)^3}{T^2} \notag \\
	\text{s.t.} \qquad &d\leq \sum_{n=1}^N D_n^T,  \label{e:tdma_first_con_thr}\\
	&2\sum_{n=1}^N t_n\leq T-\frac{Ld}{f_{B}}, \label{e:tdma_first_con_time}\\
	&0\leq d\leq D; P_n\geq 0 ~Q_n\geq 0 ~t_n\geq 0, \forall n\in \mathcal{N}.
	\end{align}
	\end{subequations}
\end{prob}

For the case of DF-FDMA, the total energy consumption of the mobile device and $N$ relay nodes is
$\sum \limits_{n=1}^N \left(P_nt+Q_nt \right) +\frac{\kappa L^3 (D-d)^3}{T^2}.$
Combining the constraints in (\ref{e:fdma_w_positive}), (\ref{e:fdma_sum_w_n}), (\ref{e:fdma_sum_data}), (\ref{e:edge_comp_freq}), and imposing the non-negative restrictions on $t$ , $P_n$, $Q_n$, $w_n$ for $n \in \mathcal{N}$, and box constraint on variable $d$, the optimization problem can be formulated as follows
\begin{prob} \label{p:fdma_first} \small
	\begin{subequations}
	\begin{align}
	\min_{\substack{d,t,\{P_n|n\in \mathcal{N}\},\{Q_n|n\in \mathcal{N}\},\{w_n|n\in \mathcal{N}\}}}
	&\sum_{n=1}^N \left(P_nt+Q_nt \right) +\frac{\kappa L^3 (D-d)^3}{T^2} \notag \\
	\text{s.t.} \qquad & d\leq \sum_{n=1}^N D_n^F, \\
	& 2t\leq T-\frac{Ld}{f_{B}}, \\
	& \sum_{n=1}^N w_i \leq W, \\
	& 0\leq d\leq D, t\geq 0;   P_n\geq 0 ~Q_n\geq 0 ~w_n\geq 0, \forall n\in \mathcal{N}.
	\end{align}
	\end{subequations}
\end{prob}

For the case of AF, the energy consumption of the mobile device and $N$ relay nodes is
$\sum_{n=1}^N \beta_n^2\left(Ph_n+\sigma^2W\right)t + Pt + \frac{\kappa L^3 (D-d)^3}{T^2}.$
Collecting the constraints (\ref{e:AF_sum_data}), (\ref{e:edge_comp_freq}), and imposing non-negative restrictions on $t$, $P$, $\beta_n$ for $n\in \mathcal{N}$, and box constraint on variable $d$, the total energy consumption minimization problem can be formulated as follows
\begin{prob} \label{p:af_first} \small
	\begin{subequations}
	\begin{align}
	\min_{\substack{d, t, P, \{\beta_n|n\in \mathcal{N}\}}}
	&  \sum_{n=1}^N \beta_n^2\left(Ph_n+\sigma^2W\right)t + Pt + \frac{\kappa L^3 (D-d)^3}{T^2} \notag \\
	\text{s.t.} \quad &d\leq tW\ln \left(1+\frac{P\left(\sum_{n=1}^{N} \sqrt{h_ng_n} \beta_n \right)^2}{\sigma^2 W \left(1+\sum_{n=1}^N g_n\beta_n^2 \right)}\right), \label{e:af_first_con_data}\\
	& 2t\leq T-\frac{Ld}{f_{B}}, \label{e:af_first_con_time} \\
	& 0\leq d\leq D, t\geq 0, P \geq 0; \beta_n\geq 0, \forall n\in \mathcal{N}.
	\end{align}
	\end{subequations}
\end{prob}

To this end, the problem formulation for the energy consumption minimization problem for the relay modes of DF-TDMA, DF-FDMA, and AF, has been completed.

\section{Optimal Solution} \label{s:prob_solution}
In this section, Problem \ref{p:tdma_first}, Problem \ref{p:fdma_first} and Problem \ref{p:af_first}, which are associated with the cases of DF-TDMA, DF-FDMA, and AF respectively, will be solved.

\subsection{Solution for the Case of DF-TDMA} \label{s:tdma_solutoin}

To simplify the solving of Problem \ref{p:tdma_first}, the following lemma is claimed first.
\begin{lem} \label{lem:tdma_p_q_equal}
For Problem \ref{p:tdma_first}, the optimal $P_n$ and $Q_n$ for $n\in \mathcal{N}$ should satisfy
\begin{equation} \label{e:tdma_power_and_gain}
	P_n h_n=Q_n g_n, \forall n \in \mathcal{N}.
\end{equation}
\end{lem}
\begin{IEEEproof}
Please refer to Appendix \ref{A:lem_tdma_p_q_equal}.
\end{IEEEproof}
With Lemma \ref{lem:tdma_p_q_equal}, by expressing $Q_n$ with $P_n$ acoording to (\ref{e:tdma_power_and_gain}) for the objective function of Problem \ref{p:tdma_first} and constraint (\ref{e:tdma_first_con_thr}), Problem \ref{p:tdma_first} turns to be the following optimization problem
\begin{prob} \label{p:tdma_deleteq} \small
	\begin{subequations}
	\begin{align}
	\min_{\substack{d, \{t_n|n\in \mathcal{N}\}, \{P_n|n\in \mathcal{N}\}}}
	&\sum_{n=1}^N P_nt_n \left(1+\frac{h_n}{g_n} \right) +\frac{\kappa L^3 (D-d)^3}{T^2} \notag \\
	\text{s.t.} \quad &d\leq \sum_{n=1}^N t_nW\ln \left(1+\frac{P_nh_n}{\sigma^2W}\right), \label{e:tdma_deleteq_con_data} \\
	& 2\sum_{n=1}^N t_n\leq T-\frac{Ld}{f_{B}}, \label{e:tdma_deleteq_con_time} \\
	& 0\leq d\leq D; P_n\geq 0 ~t_n\geq 0, \forall n\in \mathcal{N}.
	\end{align}
	\end{subequations}
\end{prob}

Problem \ref{p:tdma_deleteq} is non-convex due to the non-convexity of the function in the objective function of Problem \ref{p:tdma_deleteq} and the non-concavity of the right-hand side function of the constraint (\ref{e:tdma_deleteq_con_data}) with variables $P_n$ and $t_n$ for $n\in \mathcal{N}$.
To find the optimal solution, define $E_n = P_n t_n$ and express $P_n$ with $E_n$ and $t_n$, then Problem \ref{p:tdma_deleteq} is equivalent with the following optimization problem
\begin{prob} \label{p:tdma_convex} \small
	\begin{subequations}
	\begin{align}
	\min_{\substack{d,\{t_n|n\in \mathcal{N}\}, \{E_n|n\in \mathcal{N}\}}}
	&\sum_{n=1}^N E_n \left(1+\frac{h_n}{g_n} \right) +\frac{\kappa L^3 (D-d)^3}{T^2} \notag\\
	\text{s.t.} \quad &d\leq \sum_{n=1}^N t_nW\ln \left(1+\frac{E_nh_n}{t_n\sigma^2W} \right), \label{e:tdma_convex_con_data} \\
	&2\sum_{n=1}^N t_n\leq T-\frac{Ld}{f_{B}}, \label{e:tdma_convex_con_time} \\
	&0\leq d\leq D; E_n\geq 0 ~t_n\geq 0, \forall n\in \mathcal{N}.
 	\end{align}
 	\end{subequations}
\end{prob}
It can be checked that Problem \ref{p:tdma_convex} is a convex problem, due to the convexity of its objective function with $E_n$ for $n\in \mathcal{N}$ and $d$, and the concavity of right-hand side function of constraint (\ref{e:tdma_convex_con_data}) with $\left(E_n, t_n\right)^T$.
Hence Problem \ref{p:tdma_convex} can be solved optimally by existing algorithms, such as interior-point method \cite{convex_book}.
To further reduce the computation complexity, we offer a different solution method for Problem \ref{p:tdma_convex} by decomposing it into two levels.

In the lower level, $d$ is fixed and the following optimization problem need to be solved.
\begin{prob} \label{p:tdma_lower} \small
	\begin{subequations}
	\begin{align}
 	U(d) \triangleq
 	\min_{\substack{\{E_n|n\in \mathcal{N}\}, \{t_n|n\in \mathcal{N}\}}}
	&\sum_{n=1}^N E_n \left(1+\frac{h_n}{g_n} \right) \notag \\
	\text{s.t.} \quad &d\leq \sum_{n=1}^N t_nW\ln \left(1+\frac{E_nh_n}{t_n\sigma^2W} \right), \label{e:tdma_lower_con_data}\\
	& 2\sum_{n=1}^N t_n\leq T-\frac{Ld}{f_{B}}, \label{e:tdma_lower_con_time}\\
	& E_n\geq 0 ~t_n\geq 0, \forall n\in \mathcal{N}.
	\end{align}
	\end{subequations}
\end{prob}
In the upper level, the following optimization problem, which is equivalent with Problem \ref{p:tdma_convex}, needs to be solved 
\begin{prob} \label{p:tdma_upper} \small
	\begin{subequations}
	\begin{align}
	\min_d \quad &U(d)+\frac{\kappa L^3 (D-d)^3}{T^2} \notag \\
	\text{s.t.} \quad &0\leq d\leq D.
	\end{align}
	\end{subequations}
\end{prob}

It can be checked that Problem \ref{p:tdma_lower} is convex and satisfies Slater's conditions \cite{convex_book}. Then the KKT conditions, which can serve as sufficient and necessary condition for its optimal solution, are listed below
\begin{subequations} \small
	\small
	\begin{align}
	1+\frac{h_n}{g_n} -\frac{\mu t_nh_nW}{t_n\sigma^2W+E_nh_n} -\zeta_n=0, ~\forall n\in \mathcal{N} \label{e:tdma_KKT_dev_E} \\
	2\lambda -\mu W\left(\ln \left(1+\frac{E_nh_n}{t_n\sigma^2W} \right) -\frac{\frac{E_nh_n}{t_n\sigma^2W} }{1+\frac{E_nh_n}{t_n\sigma^2W} } \right) -\eta_n=0,
	~\forall n\in \mathcal{N} \label{e:tdma_KKT_dev_t} \\
	\mu \left(d- \sum_{n=1}^N t_nW\ln \left(1+\frac{E_nh_n}{t_n\sigma^2W} \right) \right) =0 \\
	\lambda \left(2\sum_{n=1}^N t_n+\frac{Ld}{f_B} -T \right) =0 \\
	\zeta_n E_n=0, ~\forall n\in \mathcal{N} \\
	\eta_n t_n=0, ~\forall n\in \mathcal{N} \label{e:tdma_KKT_eta_multiply} \\
	d\leq \sum_{n=1}^N t_nW\ln \left(1+\frac{E_nh_n}{t_n\sigma^2W} \right) \label{e:tdma_KKT_con_data} \\
	2\sum_{n=1}^N t_n\leq T-\frac{Ld}{f_B} \label{e:tdma_KKT_con_time} \\
	E_n\geq 0, ~\forall n\in \mathcal{N} \label{e:tdma_KKT_con_E} \\
	t_n\geq 0, ~\forall n\in \mathcal{N} \label{e:tdma_KKT_con_t}
	\end{align}
\end{subequations}
in which $\mu$, $\lambda$, $\zeta_n$ and $\eta_n$ are non-negative Lagrange multipliers associated with constraints (\ref{e:tdma_KKT_con_data}), (\ref{e:tdma_KKT_con_time}), (\ref{e:tdma_KKT_con_E}) and (\ref{e:tdma_KKT_con_t}), respectively.

Define the index set $\mathcal{P}_T=\{n| E_n > 0, t_n >0, n \in \mathcal{N} \}$ and $\bar{\mathcal{P}}_T = \{n | E_n =0, t_n =0, n \in \mathcal{N} \}$.
For the optimal solution of Problem \ref{p:tdma_lower}, it is meaningless to set $E_n=0$ while setting $t_n>0$ or set $t_n = 0$ while setting $E_n>0$, $\forall n \in \mathcal{N}$.
Hence $\forall n \in \mathcal{N}$, only two cases will be considered: 1) The case that $E_n = 0$ and $t_n=0$; 2) The case that $E_n >0$ and $t_n>0$.
Therefore, there is $\mathcal{P}_T \bigcup \bar{\mathcal{P}}_T = \mathcal{N}$.
Define $\text{SNR}_n^T=\frac{E_nh_n}{t_n\sigma^2W}$, with actually represents the signal-to-noise ratio at relay $n$, for $n\in \mathcal{N}$.
For $\text{SNR}_n^T$, $\forall n\in \mathcal{P}_T$, the following lemma can be expected.
\begin{lem} \label{lem:tdma_snr}
	For $\forall i,j\in \mathcal{P}_T$, there is
	$\text{SNR}_i^T=\text{SNR}_j^T=\text{SNR}^T$
where $\text{SNR}^T$ is a common value.
\end{lem}
\begin{IEEEproof}
Please refer to Appendix \ref{A:lem_tdma_snr}.
\end{IEEEproof}
In addition, for the optimal solution of Problem \ref{p:tdma_lower}, we have Lemma \ref{lem:tdma_equal}.
\begin{lem} \label{lem:tdma_equal}
	For optimal solution of Problem \ref{p:tdma_lower}, the equality of constraints (\ref{e:tdma_lower_con_data}) and (\ref{e:tdma_lower_con_time}) hold.
\end{lem}
\begin{IEEEproof}
	Please refer to Appendix \ref{A:lem_tdma_equal}.
\end{IEEEproof}

With the aid of Lemma \ref{lem:tdma_snr} and Lemma \ref{lem:tdma_equal}, for $\forall n \in \mathcal{P}_T$, there is
\begin{equation} \label{e:tdma_SNR_exp} \small
\text{SNR}_n^T = \text{SNR}^T
=  \frac{\sum_{n \in \mathcal{P}}E_n h_n}{\sum_{n \in \mathcal{P}} t_n \sigma^2 W }
= \frac{\sum_{n \in \mathcal{P}}E_n h_n + \sum_{n \in \bar{\mathcal{P}}}E_n h_n}{\sum_{n \in \mathcal{P}} t_n \sigma^2 W  + \sum_{n \in \bar{\mathcal{P}}} t_n \sigma^2 W}
=  \frac{\sum_{n \in \mathcal{N}}E_n h_n}{\sum_{n \in \mathcal{N}} t_n \sigma^2 W }
= \frac{2\sum_{n \in \mathcal{N}}E_n h_n}{\left(T - \frac{Ld}{f_B}\right)\sigma^2 W},
\end{equation}
which further indicates
\begin{equation} \small
t_n = \frac{ E_n h_n}{\sigma^2 W \text{SNR}^T} = \frac{E_n h_n \left(T - \frac{Ld}{f_B}\right)}{2 \sum_{n \in \mathcal{N}} E_n h_n}, \forall n \in \mathcal{P}_T.
\end{equation}
In addition, there naturally exists
\begin{equation} \small
t_n = \frac{E_n h_n \left(T - \frac{Ld}{f_B}\right)}{2 \sum_{n \in \mathcal{N}} E_n h_n}, \forall n \in \bar{\mathcal{P}}_T.
\end{equation}
Hence the solution of $t_n$ can be expressed by $E_n$ for $n\in \mathcal{N}$ and we only need to focus on finding the optimal solution of $E_n$ for $n \in \mathcal{N}$.

By combing Lemma \ref{lem:tdma_equal} and (\ref{e:tdma_SNR_exp}), the (\ref{e:tdma_lower_con_data}) can be simplified to be
\begin{equation} \small
\sum_{n=1}^N E_nh_n\geq \frac{1}{2} \sigma^2W\left(T-\frac{Ld}{f_B} \right) \left(e^{\frac{2d}{W\left(T-\frac{Ld}{f_B} \right)}}-1 \right).
\end{equation}
Then Problem \ref{p:tdma_lower} is equivalent with the following optimization problem
\begin{prob} \label{p:tdma_lower_final} \small
	\begin{subequations}
	\begin{align}
	U(d) = \min_{\{E_n|n \in \mathcal{N}\}}
	&\sum_{n=1}^N E_n \left(1+\frac{h_n}{g_n} \right) \notag \\
	\text{s.t.} \quad &\sum_{n=1}^N E_nh_n\geq \frac{1}{2} \sigma^2W\left(T-\frac{Ld}{f_B} \right)
	\left(e^{\frac{2d}{W\left(T-\frac{Ld}{f_B} \right)}}-1 \right), \label{e:tdma_lower_final_con_data}\\
	&E_n\geq 0, \forall n\in \mathcal{N}.
	\end{align}
	\end{subequations}
\end{prob}
Problem \ref{p:tdma_lower_final} is a linear programming problem with respect to the vector of $\{E_n|n\in \mathcal{N}\}$ and can be solved by existing numerical methods, e.g., simplex method and interior-point method \cite{numerical_book}.

Next, we turn to solve the upper level problem, i.e., Problem \ref{p:tdma_upper}, by optimizing the variable $d$. For Problem \ref{p:tdma_upper}, Lemma \ref{lem:tdma_upper_convex} establishes.
\begin{lem} \label{lem:tdma_upper_convex}
	Problem \ref{p:tdma_upper} is a convex optimization problem.
\end{lem}
\begin{IEEEproof}
	Please refer to Appendix \ref{A:lem_tdma_upper_convex}.
\end{IEEEproof}	
According to Lemma \ref{lem:tdma_upper_convex}, golden-search method can be utilized to find the optimal solution of Problem \ref{p:tdma_upper}.


\subsection{Solution for the Case of DF-FDMA} \label{s:fdma_solution}

In this case, by following the similar discussion in Lemma \ref{lem:tdma_p_q_equal}, Problem \ref{p:fdma_first} turns to be
\begin{prob} \label{p:fdma_deleteq} \small
	\begin{subequations}
	\begin{align}
	\min_{\substack{d,t,\{P_n|n\in \mathcal{N}\}, \{w_n|n\in \mathcal{N}\}}}
	&\sum_{n=1}^N P_nt \left(1+\frac{h_n}{g_n} \right) +\frac{\kappa L^3 (D-d)^3}{T^2} \notag \\
	\text{s.t.} \qquad &d\leq \sum_{n=1}^N tw_n\ln\left(1+\frac{P_nh_n}{\sigma^2w_n} \right), \label{e:fdma_deleteq_con_data} \\
	&2t\leq T-\frac{Ld}{f_B}, \label{e:fdma_deleteq_con_time} \\
	&\sum_{n=1}^N w_n \leq W, \label{e:fdma_deleteq_con_band} \\
	& 0\leq d\leq D, t\geq 0; P_n\geq 0 ~w_n\geq 0, \forall n\in \mathcal{N}.
	\end{align}
	\end{subequations}
\end{prob}

In Problem \ref{p:fdma_deleteq}, the objective function and constraint (\ref{e:fdma_deleteq_con_data}) are non-convex due to the coupling of $t$ and $P_n$, and the coupling of $t$ and $w_n$, which make it hard to solve optimally.
To explore the optimal solution of Problem \ref{p:fdma_deleteq}, the following lemma can be anticipated.
\begin{lem} \label{lem:fdma_equal}
	For the optimal solution of Problem \ref{p:fdma_deleteq}, the equality of constraints (\ref{e:fdma_deleteq_con_data}), (\ref{e:fdma_deleteq_con_time}) and (\ref{e:fdma_deleteq_con_band}) hold.
\end{lem}
\begin{IEEEproof}
	Please refer to Appendix \ref{A:lem_fdma_equal}.
\end{IEEEproof}
With Lemma \ref{lem:fdma_equal}, we can express $t$ with $\frac{\left(Tf_B - Ld\right)}{2f_B}$.
Furthermore, the following lemma can be expected.
\begin{lem} \label{lem:fdma_problem_equivalence}
Mathematically, Problem \ref{p:tdma_convex} is equivalent with Problem \ref{p:fdma_deleteq}.
\end{lem}
\begin{IEEEproof}
	\label{A:lem_fdma_problem_equivalence}
To prove this lemma, first we make some transformation for Problem \ref{p:tdma_convex}. Define $r_n$ as $r_n=t_nW$ for $n\in \mathcal{N}$. Apply the variable substitution and utilize Lemma \ref{lem:tdma_equal}, Problem \ref{p:tdma_convex} is equivalent with the following optimization problem
\begin{prob} \label{p:tdma_equivalence} \small
	\begin{subequations}
		\begin{align}
		\min_{\substack{d,\{r_n|n\in \mathcal{N}\}, \{E_n|n\in \mathcal{N}\}}}
		&\sum_{n=1}^N E_n \left(1+\frac{h_n}{g_n} \right) +\frac{\kappa L^3 (D-d)^3}{T^2} \notag\\
		\text{s.t.} \quad &d= \sum_{n=1}^N r_n\ln \left(1+\frac{E_nh_n}{\sigma^2r_n} \right), \label{e:tdma_equivalence_con_data}\\
		&2\sum_{n=1}^N r_n= W \left(T-\frac{Ld}{f_{B}}\right), \label{e:tdma_equivalence_con_time}\\
		&0\leq d\leq D; E_n\geq 0 ~r_n\geq 0, \forall n\in \mathcal{N},
		\end{align}
	\end{subequations}
\end{prob}
in which the expression of (\ref{e:tdma_equivalence_con_data}) is straightforward and the expression of (\ref{e:tdma_equivalence_con_time}) is the result of $2\sum_{n=1}^N t_n= T-\frac{Ld}{f_{B}}$ multiplied by $W$ on both sides.

Next we transform Problem \ref{p:fdma_deleteq}.
To this end, we apply variable substitution $E_n^{\dagger}=P_nt$ and $r_n^{\dagger}=w_nt$ for $n\in \mathcal{N}$.
To avoid abuse of variables, we also replace $d$ with $d^{\dagger}$ for FDMA case.
Based on Lemma \ref{lem:fdma_equal}, constraint (\ref{e:fdma_deleteq_con_data}) is transformed into $d^{\dagger}= \sum_{n=1}^N r_n^{\dagger}\ln\left(1+\frac{E_n^{\dagger}h_n}{\sigma^2r_n^{\dagger}} \right)$.
In Problem \ref{p:fdma_deleteq}, multiplied by the equality $2t= T-\frac{Ld^{\dagger}}{f_B}$ on both sides, constraint (\ref{e:fdma_deleteq_con_band}) turns to be $2\sum_{n=1}^N r_n^{\dagger} \leq W(T-\frac{Ld^{\dagger}}{f_B})$. Furthermore, by utilizing the equation $\sum_{n=1}^N w_n = W$ claimed by Lemma \ref{lem:fdma_equal}, constraint (\ref{e:fdma_deleteq_con_time}) and constraint (\ref{e:fdma_deleteq_con_band}) can be combined to the single constraint $2\sum_{n=1}^N r_n^{\dagger} = W(T-\frac{Ld^{\dagger}}{f_B})$. Consequently, Problem \ref{p:fdma_deleteq} can be rewritten as the following optimization problem
\begin{prob} \label{p:fdma_equivalence} \small
	\begin{subequations}
		\begin{align}
		\min_{\substack{d^{\dagger},\{r_n^{\dagger}|n\in \mathcal{N}\}, \{E_n^{\dagger}|n\in \mathcal{N}\}}}
		&\sum_{n=1}^N E_n^{\dagger} \left(1+\frac{h_n}{g_n} \right) +\frac{\kappa L^3 (D-d^{\dagger})^3}{T^2} \notag \\
		\text{s.t.} \qquad &d^{\dagger}= \sum_{n=1}^N r_n^{\dagger}\ln\left(1+\frac{E_n^{\dagger}h_n}{\sigma^2r_n^{\dagger}} \right), \label{e:fdma_equivalence_con_data} \\
		&2\sum_{n=1}^N r_n^{\dagger} = W \left(T-\frac{Ld^{\dagger}}{f_B} \right), \label{e:fdma_equivalence_con_band}\\
		& 0\leq d^{\dagger}\leq D, E_n^{\dagger}\geq 0  ~r_n^{\dagger}\geq 0, \forall n\in \mathcal{N}.
		\end{align}
	\end{subequations}
\end{prob}

It can be easily observed that Problem \ref{p:tdma_equivalence} and Problem \ref{p:fdma_equivalence} are with the identical mathematical form.
This completes the proof.
\end{IEEEproof}

\begin{rem} \label{rem:fdma_problem_equivalence}
	Retrospecting the proof of Lemma \ref{lem:fdma_problem_equivalence}, the variables $d$, $E_n$ and $r_n$ in Problem \ref{p:tdma_equivalence} are corresponded with the variables $d^{\dag}$, $E_n^{\dag}$, and $r_n^{\dag}$ in Problem \ref{p:fdma_equivalence} for $n\in \mathcal{N}$, respectively.
	Hence, with a configuration of $d$, $E_n$ and $r_n$ for $n\in \mathcal{N}$ in the mode of TDMA, there always a set of
	$d^{\dag}$, $E_n^{\dag}$, and $r_n^{\dag}$ for $n\in \mathcal{N}$, such that
	\begin{equation} \label{e:tdma_fdma_equal}
	E_n=E_n^{\dagger},
	r_n=r_n^{\dagger},
	d=d^{\dagger},
	\end{equation}
	to achieve the identical total energy consumption of the mobile device and multiple relay nodes, in the mode of FDMA.
	Especially, the minimal objective function of Problem \ref{p:tdma_equivalence} is also identical with the minimal objective function of Problem \ref{p:fdma_equivalence}, which indicates that TDMA mode and FDMA mode can achieve the same minimal total energy consumption.
\end{rem}

With Lemma \ref{lem:fdma_problem_equivalence} and Remark \ref{rem:fdma_problem_equivalence}, Problem \ref{p:fdma_deleteq} can be solved optimally by following the similar way in Section \ref{s:tdma_solutoin}. To be specific, the optimal $d$ of Problem \ref{p:fdma_deleteq} is also the optimal $d$ of Problem \ref{p:tdma_convex}; the optimal $t$ of Problem \ref{p:fdma_deleteq} is $t=\frac{\left(Tf_B - Ld \right)}{2f_B}$ according to Lemma \ref{lem:fdma_equal}; the optimal $P_n$ of Problem  \ref{p:fdma_deleteq} for $n\in \mathcal{N}$ is $P_n = \frac{E_n^*} {t}$ where $E_n^*$ is the optimal $E_n$ of Problem \ref{p:tdma_convex}; and the optimal $w_n$ of Problem \ref{p:fdma_deleteq} for $n\in \mathcal{N}$ is $w_n = \frac{t_n^* W}{t}$ where $t_n^*$ is the optimal $t_n$ of Problem \ref{p:tdma_convex} and $t=\frac{\left(Tf_B - Ld \right)}{2f_B}$.
To this end, the optimal solution of Problem \ref{p:fdma_deleteq} is obtained.

\subsection{Solution for the Case of AF} \label{s:AF_solution}
For the case of AF, Problem \ref{p:af_first} needs to be solved. Similar with the discussion in Section \ref{s:tdma_solutoin} and Section \ref{s:fdma_solution}, the following lemma can be expected.
\begin{lem} \label{lem:af_equal}
	For optimal solution of Problem \ref{p:af_first}, the equality of constraints (\ref{e:af_first_con_data}) and (\ref{e:af_first_con_time}) hold.
\end{lem}
\begin{IEEEproof}
	The proof for Lemma \ref{lem:af_equal} is similar to the one of Lemma \ref{lem:tdma_equal} and Lemma \ref{lem:fdma_equal} and is omitted here.
\end{IEEEproof}

With Lemma \ref{lem:af_equal}, substitute the $t$ in Problem \ref{p:af_first} with $\frac{\left(Tf_B - Ld\right)}{2f_B}$, and with some mathematical manipulations, Problem \ref{p:af_first} can be equivalently transformed into the following form:
\begin{prob} \label{p:af_equal} \small
	\begin{subequations}
	\begin{align}
	\min_{d,P,\{\beta_n|n \in \mathcal{N}\}} &\frac{\left(T f_B-Ld\right)}{2f_B} \left(P+\sum_{n=1}^N \beta_n^2 \left(Ph_n+\sigma^2W\right)\right) +\frac{\kappa L^3 (D-d)^3}{T^2} \notag \\
	\text{s.t.} \qquad &\frac{P\left(\sum_{n=1}^{N} \sqrt{h_ng_n} \beta_n \right)^2}{\sigma^2W\left(1+\sum_{n=1}^N g_n\beta_n^2 \right)} \label{e:af_equal_con_data} \geq e^{\frac{2d}{W\left(T-\frac{Ld}{f_B}\right)}}-1, \\
	& 0\leq d\leq D, P\geq 0; \beta_n\geq 0, \forall n\in \mathcal{N}.	
	\end{align}
	\end{subequations}
\end{prob}

Problem \ref{p:af_equal} is still a non-convex problem. To make Problem \ref{p:af_equal} tractable, by following the similar way in Section \ref{s:tdma_solutoin} and Section \ref{s:fdma_solution}, Problem \ref{p:af_equal} is also decomposed into two levels. In the lower level, with $d$ given,
Problem \ref{p:af_lower} is required to be solved, which is given as
\begin{prob} \label{p:af_lower} \small
	\begin{subequations}
	\begin{align}
	X(d)= \min_{P,\{\beta_n|n\in \mathcal{N}\}} &
	P+\sum_{n=1}^N \beta_n^2 \left(Ph_n+\sigma^2W\right) \notag \\
	\text{s.t.} \quad &\frac{P\left(\sum_{n=1}^{N} \sqrt{h_ng_n} \beta_n \right)^2}{\sigma^2W\left(1+\sum_{n=1}^N g_n\beta_n^2 \right)} \geq e^{\frac{2d}{W\left(T-\frac{Ld}{f_B}\right)}}-1, \\
    & P \geq 0, \beta_n\geq 0, \forall n\in \mathcal{N}.
	\end{align}
	\end{subequations}
\end{prob}
In the upper level, Problem \ref{p:af_upper} should be solved by optimizing variable $d$.
\begin{prob} \label{p:af_upper} \small
	\begin{subequations}
	\begin{align}
	\min_{d} \quad &\frac{T-\frac{Ld}{f_B}}{2} X(d) +\frac{\kappa L^3 (D-d)^3}{T^2} \notag \\
	\text{s.t.} \quad &0\leq d\leq D.
	\end{align}
	\end{subequations}
\end{prob}

We first focus on solving the lower level optimization problem, i.e., Problem \ref{p:af_lower}.
By defining $\psi(d)=e^{\frac{2d}{W\left(T-\frac{Ld}{f_B}\right)}}-1$, introducing a slack variable $\delta$,
and denoting $q = \ln P$, $s = \ln \delta$, and $\alpha_n = \ln \beta_n$ for $n\in \mathcal{N}$,
Problem \ref{p:af_lower} would be equivalent with the following optimization problem
\begin{prob} \label{p:af_lower_geometric} \small
	\begin{subequations}
	\begin{align}
	Y(d)=
	\min_{q,s,\{\alpha_n|n \in \mathcal{N}\}}
	&\ln \left(e^q+\sum_{n=1}^N h_n e^{q+2\alpha_n} +\sigma^2 \sum_{n=1}^N e^{2\alpha_n} \right) \notag \\
	\text{s.t.} \qquad &\ln \left(\sum_{n=1}^N g_n e^{2\alpha_n} +1 \right) -s \leq -\ln \psi(d) -\ln \sigma^2W \label{e:af_lower_con_geometric_convex} \\
	&2\ln \left(\sum_{n=1}^N \sqrt{h_ng_n} e^{\alpha_n} \right) +q-s \geq 0 \label{e:af_lower_con_geometric_concave} \\
	& q\geq 1, s\geq 1, \alpha_n\geq 1, \forall n\in \mathcal{N}.
	\end{align}
	\end{subequations}
\end{prob}
It can be checked that $Y(d) = \ln X(d)$, i.e., the minimal cost of Problem \ref{p:af_lower_geometric} is the logarithm of the minimal cost of Problem \ref{p:af_lower}.

For Problem \ref{p:af_lower_geometric}, the objective function and the left-hand side function of constraints (\ref{e:af_lower_con_geometric_convex}) and (\ref{e:af_lower_con_geometric_concave}) are log-sum-exp functions, which is convex  according to geometric programming theory \cite{convex_book}.
However, constraint (\ref{e:af_lower_con_geometric_concave}) is in the form that a convex function is larger than 0, which does not define a convex set and leads to the non-convexity of Problem \ref{p:af_lower_geometric}.
To solve this non-convex problem, the successive convex approximation (SCA) method \cite{UAV_innerapproximation} is introduced.
Before going into details of SCA method, the optimization problem to be solved in $i$th step of iteration for SCA method is defined in Problem \ref{p:af_lower_iterative}.

In Problem \ref{p:af_lower_iterative}, $\left(q^i, s^i, \{\alpha_n^i\} \right)$ is a given and fixed point in the feasible region of Problem \ref{p:af_lower_geometric}. The left-hand side of (\ref{e:af_lower_con_iterative_concave}) is a first order approximation of the left-hand side of (\ref{e:af_lower_con_geometric_concave}), therefore (\ref{e:af_lower_con_iterative_concave}) becomes a linear constraint. Thus Problem \ref{p:af_lower_iterative} is a convex optimization problem and can be solved by traditional methods.
Specifically, the whole process to solve Problem \ref{p:af_lower_iterative} can be summarized in Algorithm~\ref{a:af}.

\begin{prob} \label{p:af_lower_iterative} \small
	\begin{subequations}
	\begin{align}
	\bar{Y}^i (d)=
	\min_{q,s,\{\alpha_n|n \in \mathcal{N}\}}
	&\ln \left(e^q+\sum_{n=1}^N h_n e^{q+2\alpha_n} +\sigma^2 \sum_{n=1}^N e^{2\alpha_n} \right) \notag \\
	\text{s.t.} \qquad &\ln \left(\sum_{n=1}^N g_n e^{2\alpha_n} +1 \right) -s \leq -\ln \psi(d) -\ln \sigma^2W \label{e:af_lower_con_iterative_convex} \\
	&\sum_{n=1}^N \frac{2\sqrt{h_ng_n}e^{\alpha_n^i}}{\sum_{n=1}^N \sqrt{h_ng_n}e^{\alpha_n^i}} (\alpha_n -\alpha_n^i) +(q-q^i) \notag \\
	&-(s-s^i) +2\ln \left(\sum_{n=1}^N \sqrt{h_ng_n} e^{\alpha_n^i} \right) +q^i-s^i \geq 0 \label{e:af_lower_con_iterative_concave} \\
	& q\geq 1, s\geq 1; \alpha_n\geq 1, \forall n\in \mathcal{N}.
	\end{align}
	\end{subequations}
\end{prob}
\begin{algorithm}[H]
	\caption{SCA procedure for solving Problem \ref{p:af_lower_geometric}}
	\begin{algorithmic}[1] \label{a:af}
		\STATE {Choose a small value $\epsilon$ as tolerance.}
		\STATE {Choose a feasible point $\left(q^0, s^0, \{\alpha_n^0\} \right)$ of Problem \ref{p:af_lower_geometric}, and denote the associated cost function of Problem \ref{p:af_lower_geometric} as $\bar{Y}^0$.}
		\STATE {Given the fixed point $\left(q^{i}, s^{i}, \{\alpha_n^{i}\} \right)$, solve Problem \ref{p:af_lower_iterative} to achieve $\bar{Y}^{i}$, and set the calculated optimal solution as $(q^{i+1}, s^{i+1}, \{\alpha_n^{i+1}\})$.} \label{ap:ite}
		\IF {$|\bar{Y}^i -\bar{Y}^{i-1}| \geq \epsilon$}
		\STATE  {Set $\left(q^{i+1}, s^{i+1}, \{\alpha_n^{i+1}\} \right)$ as the new fixed point and let $i=i+1$, go back to Step \ref{ap:ite}. }
		\ELSE
		\STATE {Quit. Output the solution $\left(q^{i+1}, s^{i+1}, \{\alpha_n^{i+1}\} \right)$, and return the most recent $\bar{Y}^{i+1}$, which is also denoted as $\bar{Y}$.}
		\ENDIF
	\end{algorithmic}
\end{algorithm}
For the convergence of Algorithm \ref{a:af}, the following lemma can be expected.
\begin{lem} \label{lem:af_convergence}
For arbitrary $\left(q^0, s^0, \{\alpha_n^0\} \right)$ in the feasible region of Problem \ref{p:af_lower_geometric}, Algorithm \ref{a:af} generates a sequence of $\bar{Y}^i$, which is decreasing and will converge to a stationary point.
\end{lem}
\begin{IEEEproof}
	Please refer to Appendix \ref{A:lem_af_convergence}.
\end{IEEEproof}
Lemma \ref{lem:af_convergence} shows that Algorithm \ref{a:af} can return a feasible yet stationary-point-achieving solution for Problem \ref{p:af_lower}. To this end, the lower level optimization problem has been solved. Next we turn to solve the upper level optimization problem, i.e., Problem \ref{p:af_upper}.

For Problem \ref{p:af_upper}, the following lemma can be expected
\begin{lem} \label{lem:af_upper_monotonic}
	$X(d)$ is a monotonic increasing function.
\end{lem}
\begin{IEEEproof}
Please refer to Appendix \ref{A:lem_af_upper_monotonic}.
\end{IEEEproof}
With Lemma \ref{lem:af_upper_monotonic}, the objective function of Problem \ref{p:af_upper} can be recognized as the difference of two monotonic increasing functions with respect to $d$, which are $\frac{T}{2}X(d)$ and $\left(\frac{Ld}{2 f_B} X(d) -\frac{\kappa L^3 (D-d)^3}{T^2}\right)$, respectively.
By introducing a new variable $\omega \geq 0$, Problem \ref{p:af_upper} is equivalent to
\begin{prob} \label{p:af_upper_difference} \small
	\begin{subequations}
	\begin{align}
	\max_{d,\omega} \quad & \left(\frac{Ld}{2 f_B} X(d) -\frac{\kappa L^3 (D-d)^3}{T^2}\right) +\omega \notag \\
	\text{s.t.} \quad &\omega+ \frac{T}{2}X(d) \leq \frac{T}{2}X(D), \\
	&0\leq d\leq D, \omega \geq 0.	
	\end{align}
	\end{subequations}
\end{prob}
The reason for the equivalence between Problem \ref{p:af_upper} and Problem \ref{p:af_upper_difference} is as follows.
Minimizing the objective function of Problem \ref{p:af_upper}, which can be written as $\frac{T}{2}X(d) - \left(\frac{Ld}{2 f_B} X(d) -\frac{\kappa L^3 (D-d)^3}{T^2}\right)$, is equivalent with maximizing $\left(\frac{Ld}{2 f_B} X(d) -\frac{\kappa L^3 (D-d)^3}{T^2}\right) - \frac{T}{2}X(d) + \frac{T}{2}X(D)$.
In addition, when the optimal solution of Problem \ref{p:af_upper_difference} is achieved, there is $\omega + \frac{T}{2} X(d) = \frac{T}{2} X(D)$ since the objective function of Problem \ref{p:af_upper_difference} is increasing with both $\omega$ and $d$. In this case, $\omega =  \frac{T}{2} X(D)- \frac{T}{2} X(d)$, and the objective function of Problem \ref{p:af_upper_difference} turns to be $\left(\frac{Ld}{2 f_B} X(d) -\frac{\kappa L^3 (D-d)^3}{T^2}\right)  - \frac{T}{2} X(d) + \frac{T}{2} X(D)$.

Problem \ref{p:af_upper_difference} is a standard {\it monotonic optimization problem}, in which the objective function is monotonic and all the constraints can be written as the form such that a monotonic function is larger than smaller than zero. For a monotonic optimization problem, $\varepsilon$-optimal solution can be found by {\it polyblock algorithm}\cite{monotonic}. The $\varepsilon$-optimal means the gap between the global optimal utility and the utility achieved by polyblock algorithm is upper bounded by any predefined value $\varepsilon>0$. Due to the limit of space, detail of the polyblock algorithm is not presented. Interested readers please refer to
\cite{monotonic}.

\begin{figure}
	\begin{center}
		\includegraphics[angle=0,width=0.35 \textwidth]{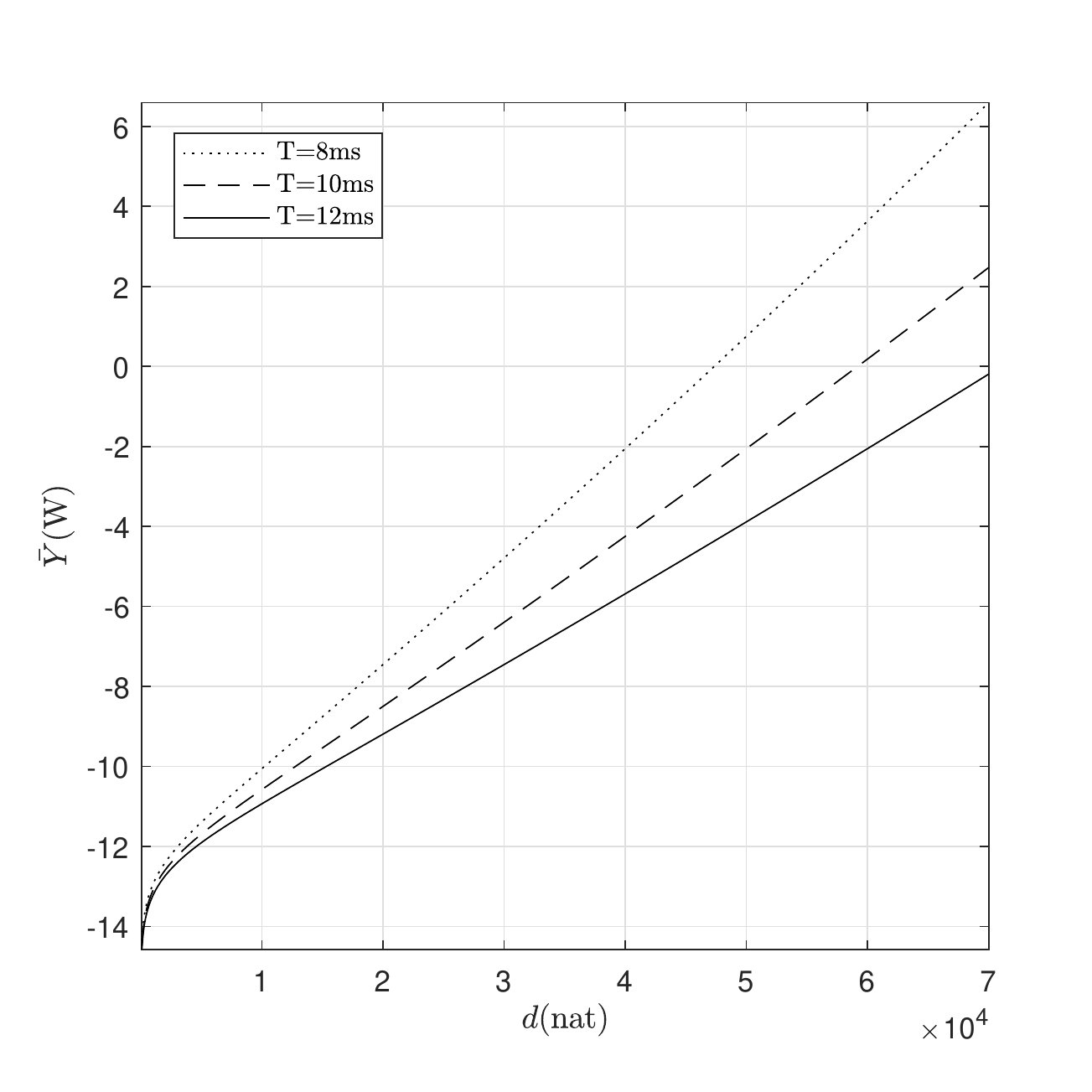}
	\end{center}
	\caption{Verification of $\bar{Y}$'s monotonicity with $d$.}
	\label{f:af_lower_monotonic}
\end{figure}

On the other hand, it should be noticed that, polyblock algorithm is utilized based on the increasing monotonicity of function $X(d)$ with $d$, or equivalently, the increasing monotonicity of $Y(d)$ with $d$. Due to the non-convexity of Problem \ref{p:af_lower_geometric}, we can only achieve a stationary point, rather than the global optimal solution, of Problem \ref{p:af_lower_geometric} by utilizing Algorithm \ref{a:af}.
In other words, we can achieve $\bar{Y}(d)$, rather than $X(d)$ or $Y(d)$.
In this case, to take the advantage of polyblock algorithm for solving the upper level problem, i.e., Problem \ref{p:af_upper}, it remains to verify that $\bar{Y}(d)$ is monotonically increasing with $d$, which, however, is very hard to prove analytically.
In Fig.\ref{f:af_lower_monotonic}, $\bar{Y}(d)$ is plotted versus $d$ under various latency requirements $T$.
In the simulation, size of the task is set to $D=7 \times 10^4$ nats, and the computation capacity $f_B=5$ GHz.
It can be seen from Fig.\ref{f:af_lower_monotonic} that $\bar{Y}(d)$ is always a monotonic increasing function with $d$. Hence we can still utilize the polyblock algorithm to find the global optimal solution if we replace $X(d)$ with $e^{\bar{Y}(d)}$ in the upper level problem, i.e., Problem \ref{p:af_upper}.

\section{Numerical Results} \label{s:numerical_results}
In this section, numerical results of our proposed algorithms for DF-TDMA mode, DF-FDMA mode and AF mode, are presented.
The default system parameters are set as follows.
The bandwidth of the whole system $W=1$ MHz. For all the relays, the distances from the mobile device to the relays and the distances from the relays to the BS are uniformly distributed between 100 meters and 500 meters.
The channels in the system experiences free space attenuation and Rayleigh fading. Hence the channel gain is the multiplication of free space path loss and the random gain under Rayleigh fading.
The free space path loss, which is denoted as $\text{PL}$ (in dB),  can be calculated from the following formula
\begin{equation*}
\text{PL}=32.4+20 \times \log \text{Distance} +20 \times \log \text{Bandwidth},
\end{equation*}
where the distance is in the unit of kilometer and the bandwidth is in the unit of MHz.
The random gain under Rayleigh fading obeys exponential distribution with mean being 0.5.
The power spectrum density of noise $\sigma^2=-140$ dBW/Hz. Similar to \cite{ref_97}, the coefficient for local computing $\kappa=10^{-25}$. The computation capacity $f_B=5$ GHz.
A computation task of data size being $8 \times 10^{4}$ nats is supposed to be finished within 0.01 second. Similar to \cite{ref_97}, $L=50$ cycles/nat. When employing the polyblock algorithm, the predefined gap $\varepsilon$ is set as $\varepsilon=10^{-5}$.

\subsection{Convergence and Optimality of Our Solution for AF Mode}

In this subsection, the convergence of SCA method to solve Problem \ref{p:af_lower} is verified, followed by one-dimension search of $d$ in Problem \ref{p:af_upper} to prove the optimality of polyblock algorithm.

\begin{figure}
\centering
\subfloat [Convergence of SCA method]{
\label{f:af_convergence}
\includegraphics[width=0.70\figwidth]{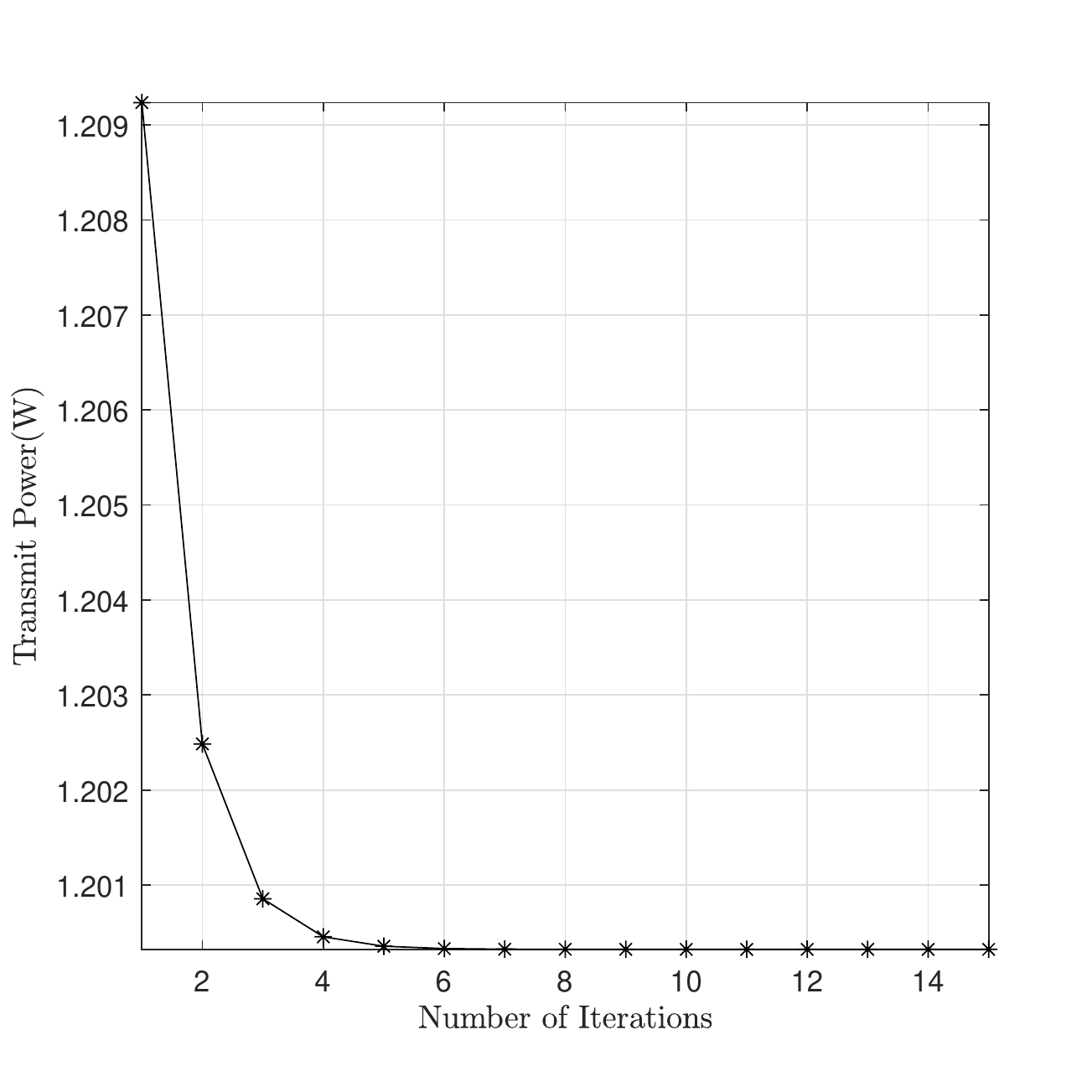}}
\subfloat [Verification of polyblock algorithm]{
\label{f:af_optimality}
\includegraphics[width=0.70\figwidth]{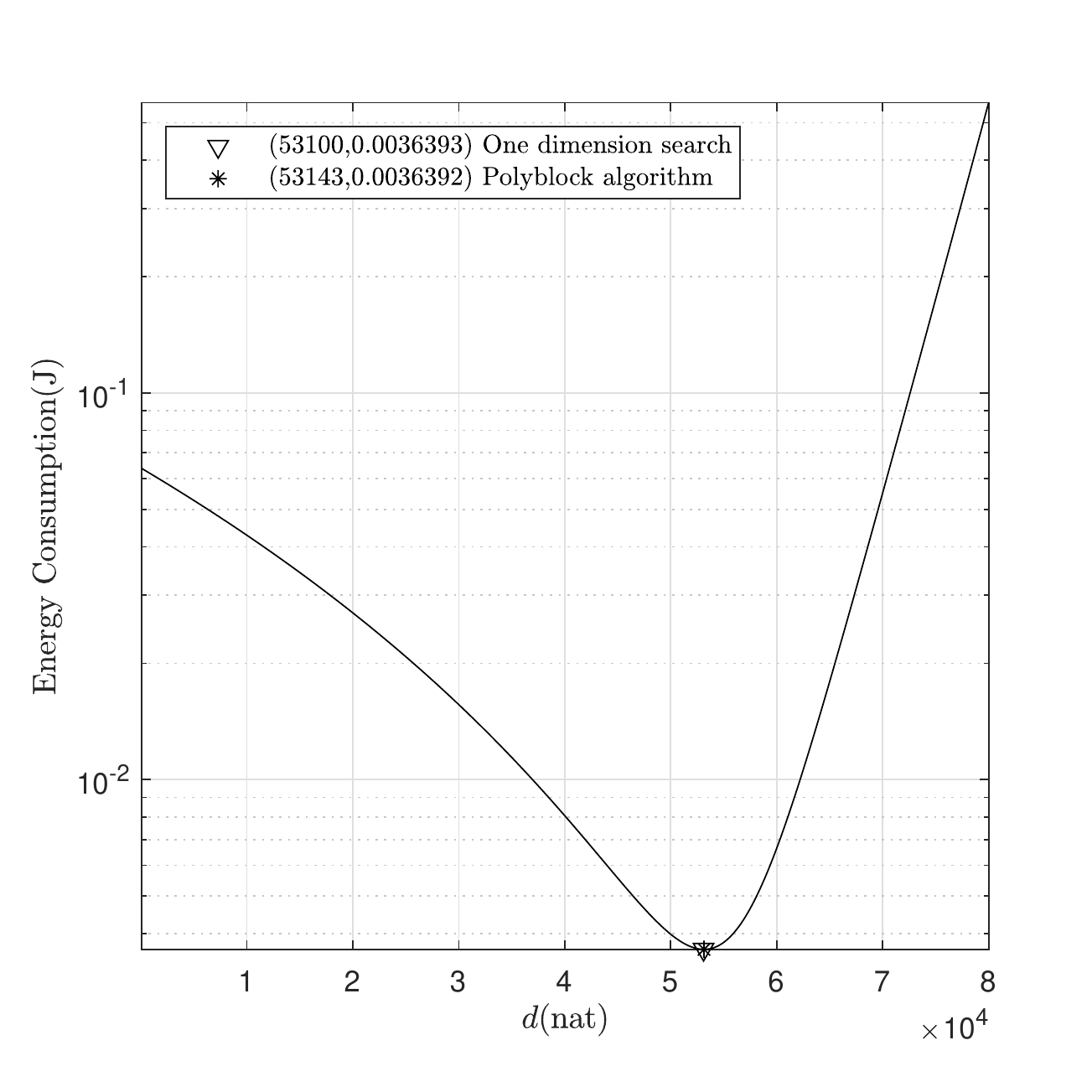}
}
\caption{Verification of convergence and optimality for AF mode.}
\end{figure}

In Fig. \ref{f:af_convergence}, the convergence of SCA method for solving Problem \ref{p:af_lower} is illustrated.
With $D=8 \times 10^4$ nats and $d=6 \times 10^4$ nats, the objective function of Problem \ref{p:af_lower} is plotted versus the number of iterations in SCA method in Fig. \ref{f:af_convergence}. By reading the plotted data points on Fig. \ref{f:af_convergence}, the objective function of Problem \ref{p:af_lower} will converge to 1.201W with a deviation of no more than $10^{-5}$ after 15 iterations.

In Fig. \ref{f:af_optimality} the objective function of Problem \ref{p:af_upper} is plotted versus $d$.
With $D=8 \times 10^4$ nats and stepsize set as 100 nats, it can be seen from Fig. \ref{f:af_optimality} that the objective function of Problem \ref{p:af_upper} is unimodel with $d$ and the minimum is achieved at $d=5.31 \times 10^4$ nats, which is very close to calculated $d$ by the polyblock algorithm. By inspecting the plotted data point and curve of Fig. \ref{f:af_optimality}, it can be also found that the gap between the minimal objective function of Problem \ref{p:af_upper} found by one-dimensional search of $d$ and the one calculated by polyblock algorithm is within the predefined $\varepsilon$.


\subsection{Complexity Analysis and Comparison of Our Proposed Method for DF-TDMA Mode}
\begin{figure}
	\begin{center}
		\includegraphics[angle=0,width=0.35 \textwidth]{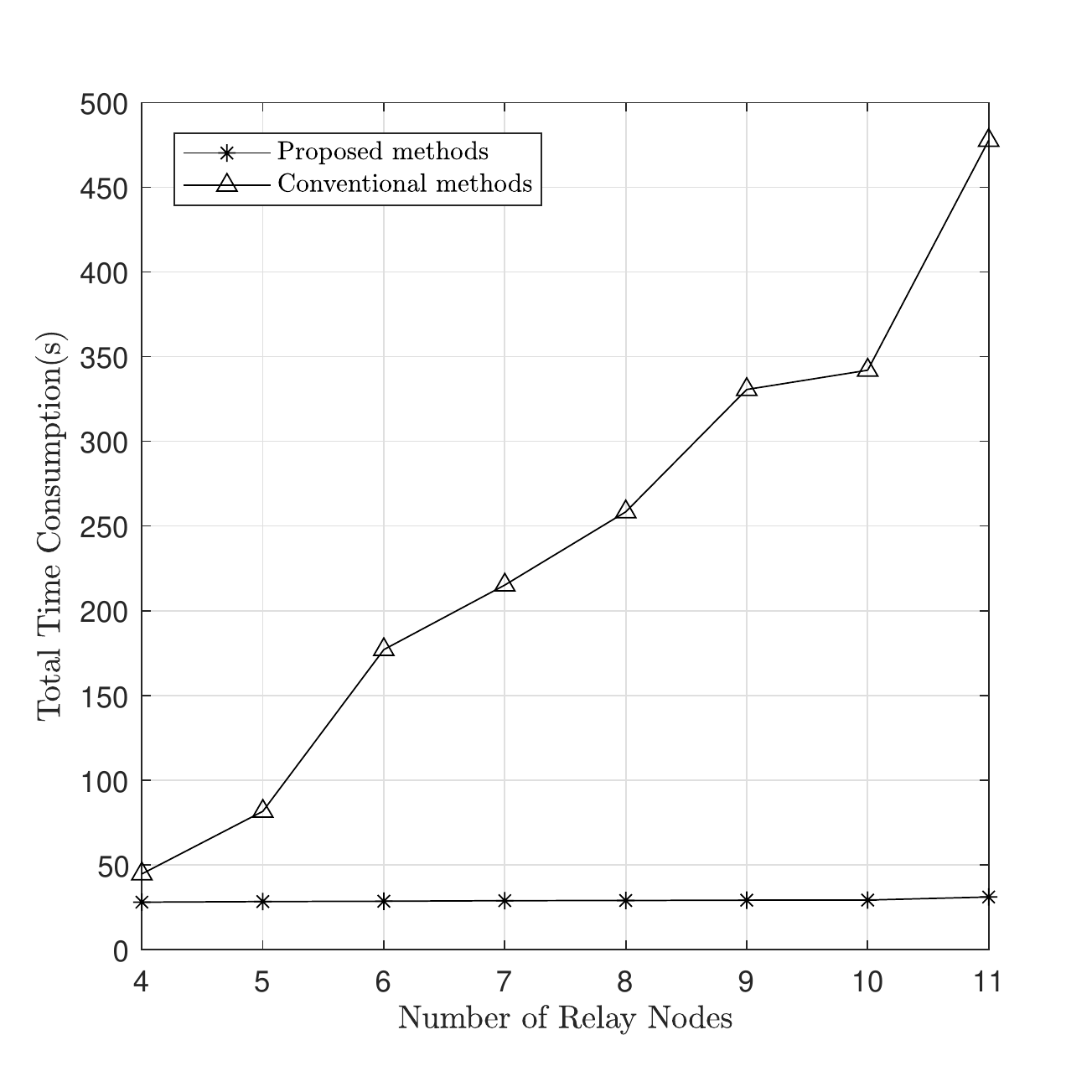}
	\end{center}
	\caption{Total time consumption of conventional and proposed methods in DF-TDMA case.}
	\label{f:time_comparison_tdma}
\end{figure}

In this subsection, the computation complexity of our proposed method for solving Problem \ref{p:tdma_convex} in DF-TDMA mode is analyzed. Note that Problem \ref{p:tdma_convex} is a convex optimization problem, which can be solved by existing algorithms optimally. As a comparison, the computation complexity for solving Problem \ref{p:tdma_convex} via  interior method, which is one of most popular numerical methods for solving a convex optimization problem optimally, is also investigated.
Fig. \ref{f:time_comparison_tdma} plots the accumulated time consumption of calculation for our proposed method and the interior-point method over 100 sets of randomly generated channel gains.
It can be seen from Fig. \ref{f:time_comparison_tdma} that as the number of relays increases, which indicates the involvement of more constraints and more variables to optimize, the total time consumption of our proposed method grows slightly while the one for interior-point method grows a lot. In addition, our proposed method can always lead to less time consumption compared with the interior-point method. This proves the advantage and necessity of our proposed method for solving Problem \ref{p:tdma_convex}.

\subsection{Performance Analysis and Comparison under DF-TDMA, DF-FDMA, and AF Modes}  \label{s:final_results}

In this subsection, two indexes, including the optimal amount of data to offload, i.e., the optimal $d$, and the minimal total energy consumption of mobile device and relays, are plotted versus the total amount of data for computing $D$, maximal tolerable delay $T$, and $f_B$ under the modes of DF-TDMA, DF-FDMA, and AF, respectively.
No method directly applicable for our investigated system can be found in existing literature.
To make comparison, two intuitive methods are also investigated for comparison, which set $t_n$ for $n\in \mathcal{N}$ equally in DF-TDMA mode, and set $w_n$ for $n\in \mathcal{N}$ equally in DF-FDMA mode.
According to Remark \ref{rem:fdma_problem_equivalence}, these two methods will achieve the same total energy consumption. For the ease of presentation, these two curves are plotted once and marked as ``equal allocation''.

\begin{figure}
\centering
\subfloat [Optimal $d$ versus $D$]{
\label{f:data_for_D}
\includegraphics[width=0.70\figwidth]{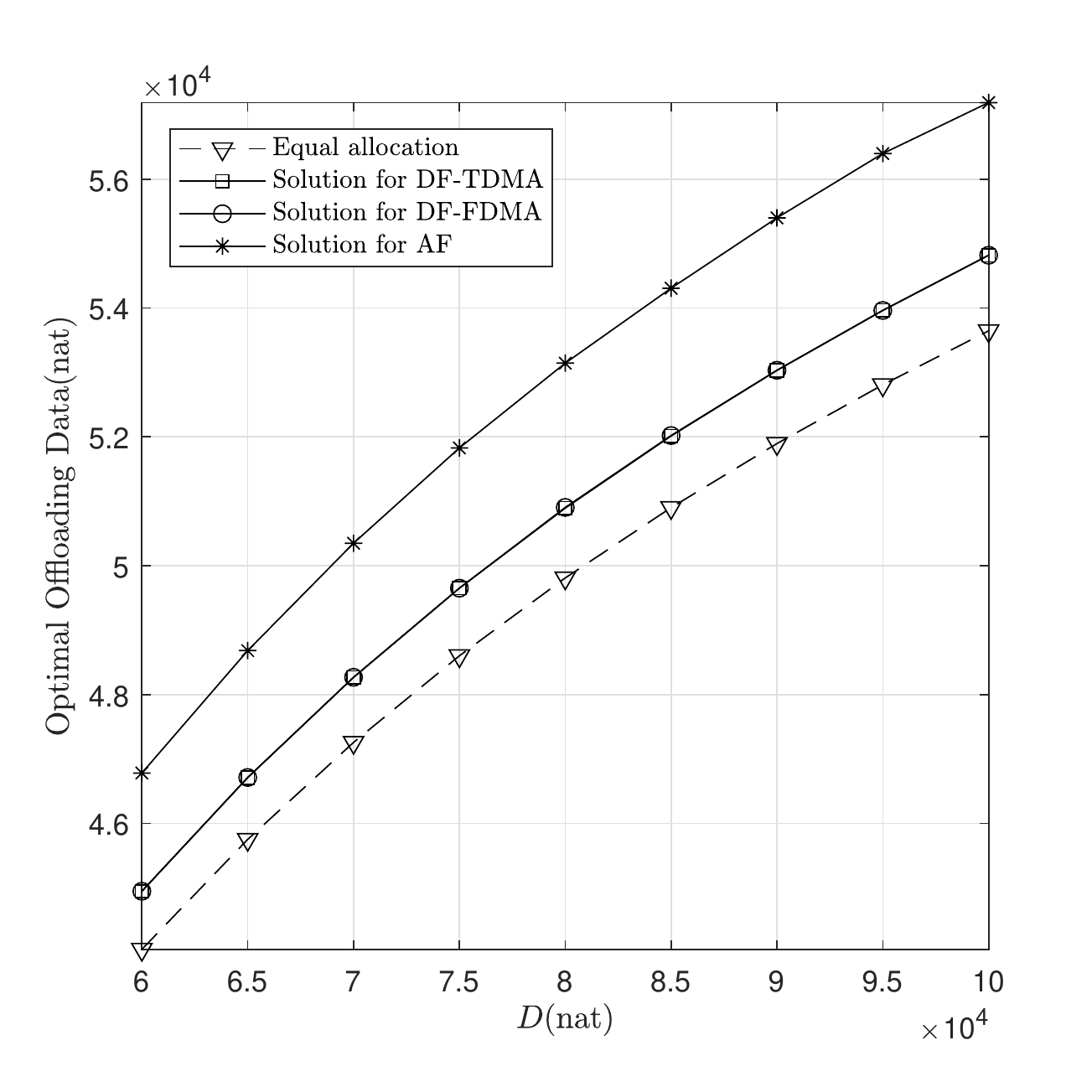}}
\subfloat [Minimal total energy consumption versus $D$]{
\label{f:energy_for_D}
\includegraphics[width=0.70\figwidth]{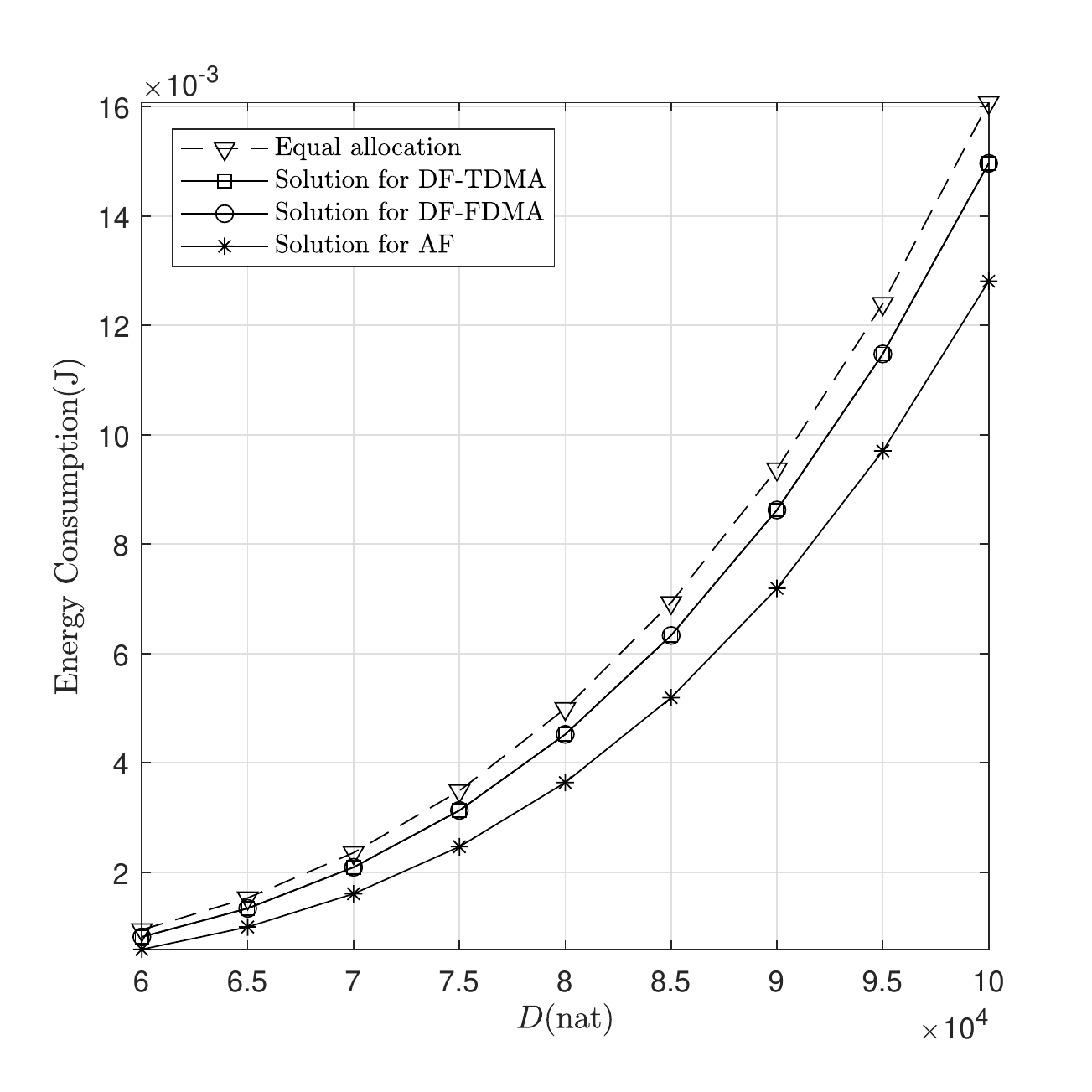}
}

\subfloat [Optimal $d$ versus $T$]{
\label{f:data_for_T}
\includegraphics[width=0.70\figwidth]{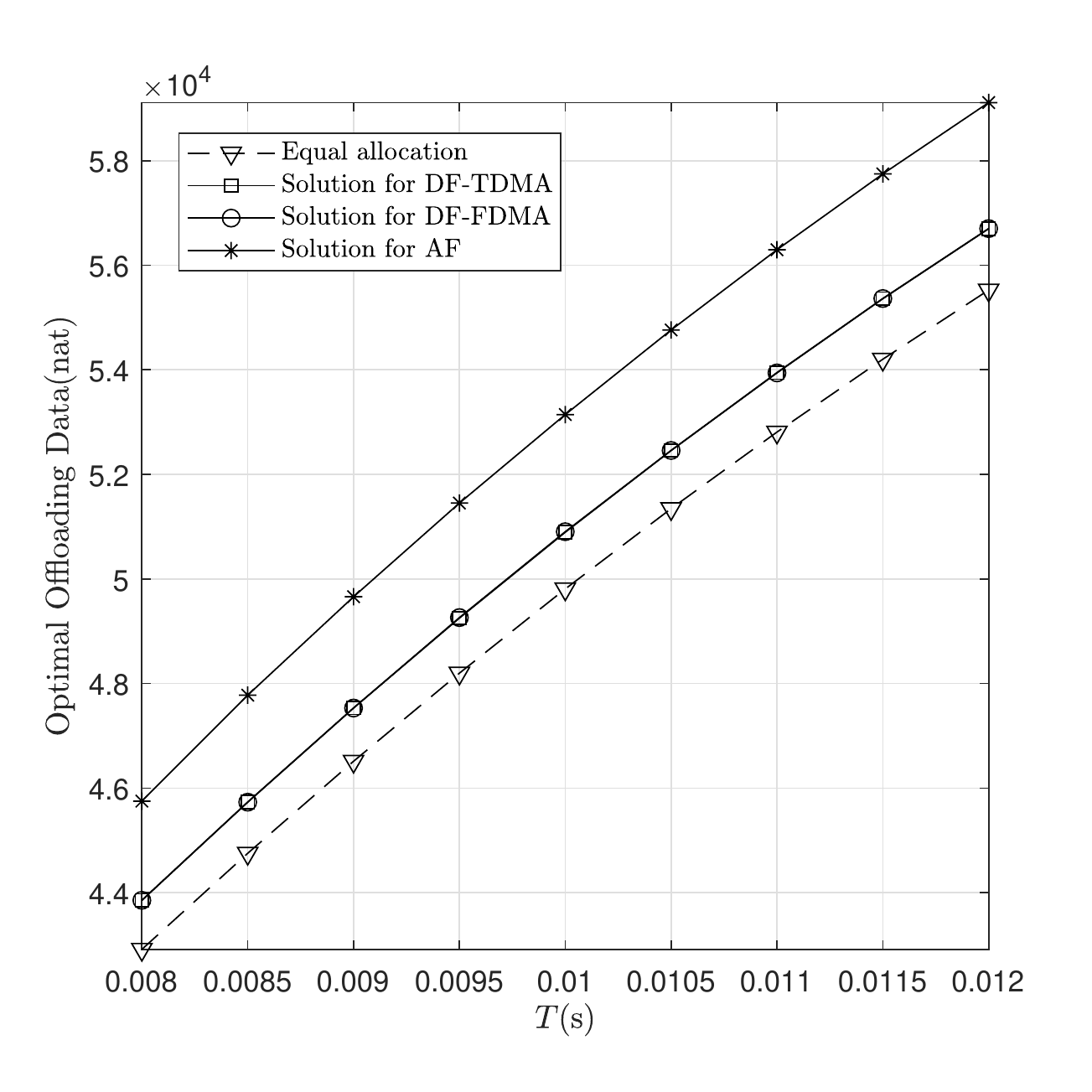}}
\subfloat [Minimal total energy consumption versus $T$]{
\label{f:energy_for_T}
\includegraphics[width=0.70\figwidth]{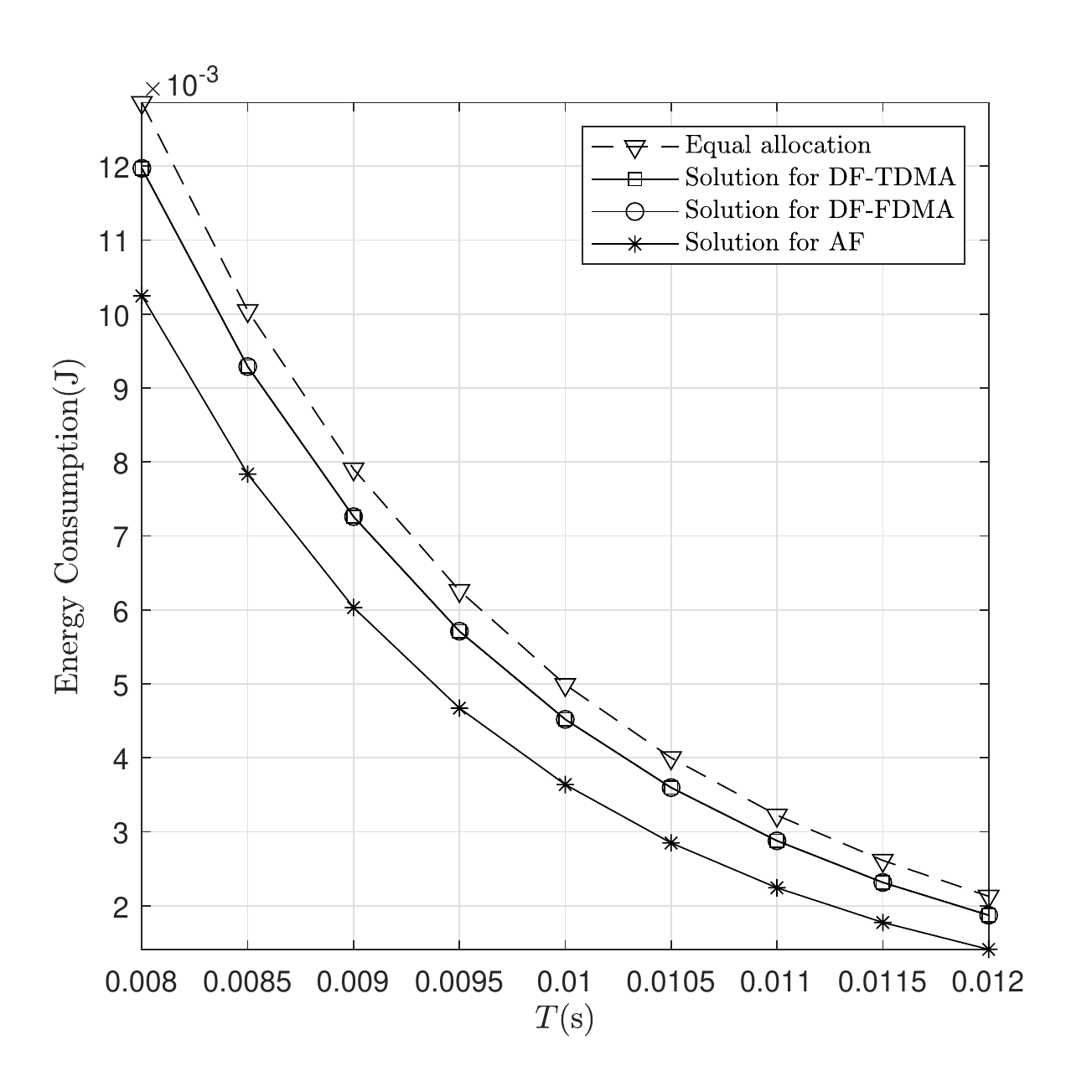}
}

\subfloat [Optimal $d$ versus $f_B$]{
\label{f:data_for_fmax}
\includegraphics[width=0.70\figwidth]{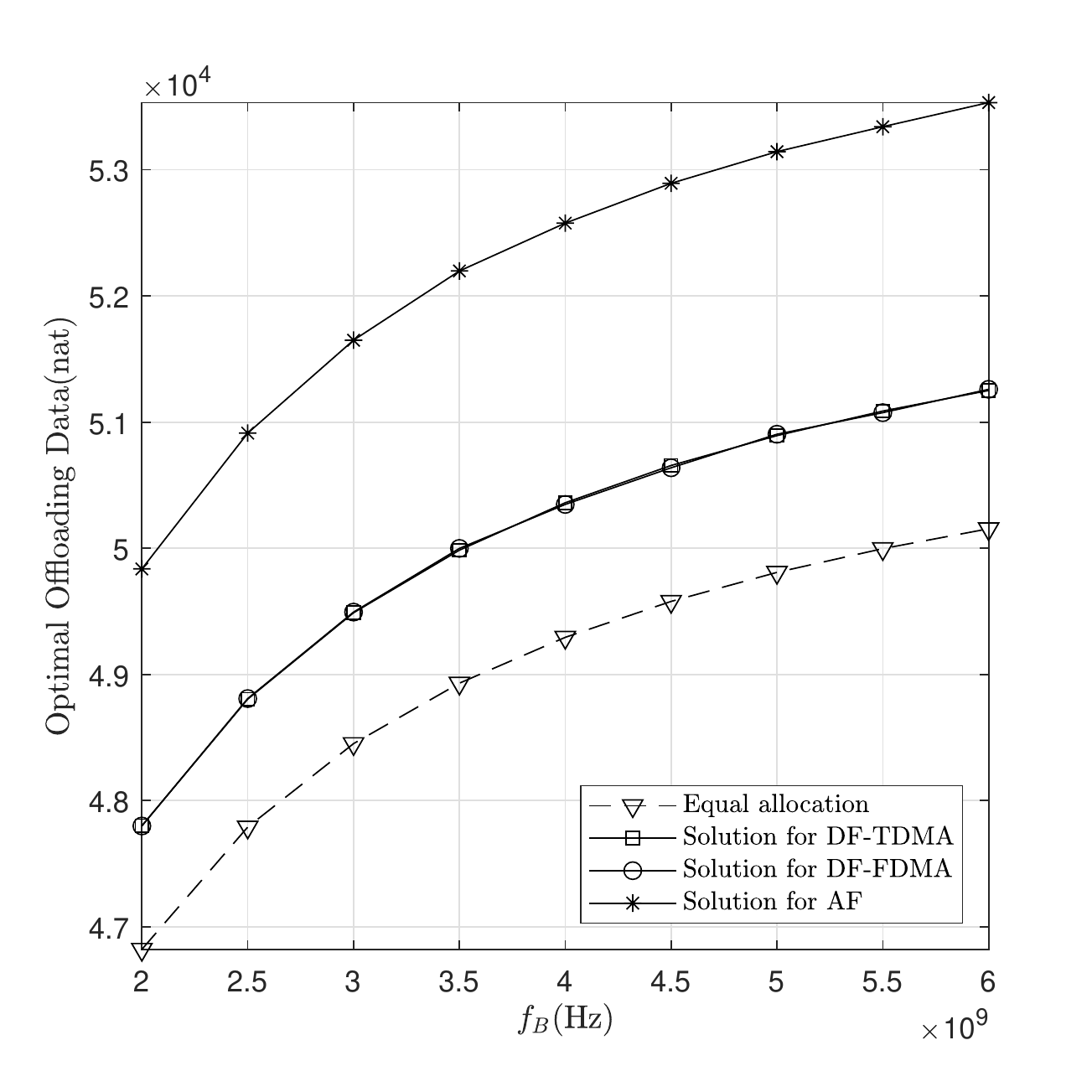}}
\subfloat [Minimal total energy consumption versus $f_B$]{
\label{f:energy_for_fmax}
\includegraphics[width=0.70\figwidth]{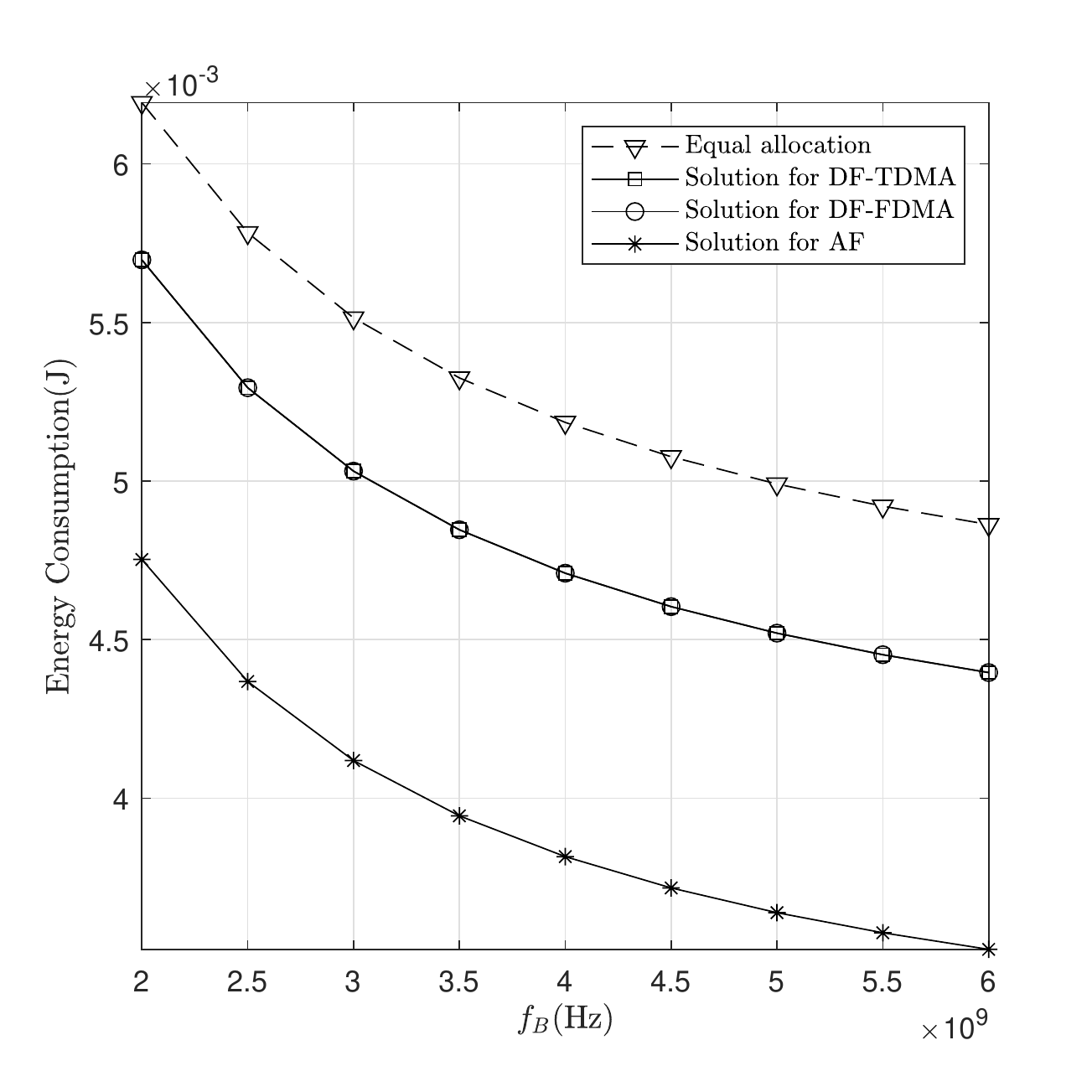}
 }

\caption{Performance analysis and comparison.}
\end{figure}



By setting $T=0.01$ second and $f_B=5$ GHz, Fig. \ref{f:data_for_D} shows the optimal amount of data to offload  versus $D$, i.e., the total amount of data for computing, and Fig. \ref{f:energy_for_D} plots the minimal total energy consumption versus $D$.
It can be seen that as $D$ grows, both the optimal amount of data to offload and the minimal total energy consumption will increase. This is in coordination with our intuition: With more $D$ to process by the MEC system, more energy consumption will be incurred naturally, and more amount of data would be offloaded to the BS given that the existence of BS can always help the mobile device to reduce energy consumption.


With $D=8 \times 10^4$ nats and $f_B=5$ GHz, Fig. \ref{f:data_for_T} and Fig. \ref{f:energy_for_T} plot the optimal amount of data to offload and the minimal total energy consumption versus maximal tolerable delay $T$, respectively.
It can be checked from Fig. \ref{f:energy_for_T} that as $T$ grows  the minimal total energy consumption will decrease. This is because the enlarged $T$ contributes to not only the relaxed feasible region but also the decreased energy consumption at local in the objective function for the associated optimization problems in various working modes.
It can be also seen from Fig. \ref{f:data_for_T} that the amount of data to offload will decrease with $T$.
From this figure we can find that it is still preferable to offload more data to the BS although enlarged $T$ can decrease the energy consumption of local computing given that the BS can saving energy consumption from computing at local.


By setting $D=8 \times 10^4$ nats and $T=0.01$ second, Fig. \ref{f:data_for_fmax} and Fig. \ref{f:energy_for_fmax} show the optimal amount of data to offload and the minimal total energy consumption versus computation capacity at the BS $f_{B}$.
It can be found from Fig. \ref{f:data_for_fmax} and Fig. \ref{f:energy_for_fmax} that the optimal amount of data to offload and the total energy consumption will go with $f_B$ by following the similar trend with $T$.
With the growth of $f_B$, the feasible region of associated optimization for any working mode is enlarged, which leads to the decrease of minimal total energy consumption. In addition, the growth of $f_B$ enables the BS to complete the computing of the same amount of data within less time, which is more attractive for the mobile device to offload data to the BS.

At last, by comparing Fig. \ref{f:data_for_D}, Fig. \ref{f:energy_for_D}, Fig. \ref{f:data_for_T}, Fig. \ref{f:energy_for_T}, Fig. \ref{f:data_for_fmax}, and Fig. \ref{f:energy_for_fmax} together, some common results can be summarized as follows.
\begin{itemize}
       \item In terms of minimal total energy consumption, the AF mode outperforms both DF-TDMA mode and DF-FDMA mode, while DF-TDMA mode and DF-FDMA mode can achieve exactly the same performance.
This result verifies the equivalence between Problem \ref{p:tdma_convex} and Problem \ref{p:fdma_deleteq} and renders such a suggestion on the selection of working mode in a real MEC system with relays: AF mode is more preferable and there should be no preference between DF-TDMA mode and DF-FDMA mode.
	\item Our proposed methods for DF-TDMA mode and DF-FDMA mode can always lead to less energy consumption than ``equal allocation'' method. This result proves the effectiveness of our proposed methods.
\end{itemize}

\section{Conclusions} \label{s:conclusion}
In this paper, we have investigated a MEC system aided by multiple relays working in DF-TDMA mode, DF-FDMA mode, and AF mode, respectively.
For these three investigated working modes, not only the amount of data to offload and slot duration for offloading are optimized, but also the slot duration over every link are jointly optimized in DF-TDMA mode, the bandwidth and transmit power over every link are jointly optimized for DF-FDMA mode, and the total transmit power of the mobile device and amplifying coefficient on every relay are jointly optimized for AF mode.
Although being non-convex, by decomposing the associated optimization problem into two levels, simple and global optimal solution is found for DF-TDMA mode by Golden-search method, and convergent solution is found for DF-AF mode by resorting to monotonic optimization and SCA method in the upper level and lower level respectively. In addition, global optimal solution is found for DF-FDMA mode by proving the mathematical equivalence between the transformed optimization problems in DF-FDMA mode and DF-TDMA mode.
Numerical results verifies the effectiveness of our proposed method.
This research could provide helpful insight on joint computation and communication resource allocation under various working modes for a relay assisted MEC system.


\appendices

\section{Proof of Lemma \ref{lem:tdma_p_q_equal}} \label{A:lem_tdma_p_q_equal}
Suppose there is a pair of $P_n$ and $Q_n$, who are optimal for Problem \ref{p:tdma_first} but do not satisfy (\ref{e:tdma_power_and_gain}).
Without loss of generality, we assume $P_n h_n \leq Q_n g_n$.
In this case,
$t_nW \ln (1+\frac{P_n h_n}{\sigma^2W})\leq t_nW \ln (1+\frac{Q_n g_n}{\sigma^2W})$ and
$D_n^F = t_nW \ln (1+\frac{P_nh_n}{\sigma^2W})$.
By defining $Q'_n$ to be $\frac{P_n h_n}{g_n}$, it can be checked that $Q'_n g_n < P_n h_n$, which indicates that a lower value of Problem \ref{p:tdma_first}'s objective function can be achieved for $P_n$ and $Q'_n$ compared with $P_n$ and $Q_n$. This contradicts the assumption that $P_n$ and $Q_n$ are optimal. Hence $P_n$ and $Q_n$ should satisfy (\ref{e:tdma_power_and_gain}).


\section{Proof of Lemma \ref{lem:tdma_snr}} \label{A:lem_tdma_snr}
For $E_n>0$ and $t_n>0$, $\eta_n$ is zero according to (\ref{e:tdma_KKT_eta_multiply}), it can be derived from (\ref{e:tdma_KKT_dev_t}) that
	\begin{equation}
	\ln \left(1+\text{SNR}_n^T \right) -\frac{\text{SNR}_n^T}{1+\text{SNR}_n^T} =\frac{2\lambda}{\mu W}, ~\forall n\in \mathcal{N}.
	\end{equation}
Looking into the function $\theta (x)=\ln \left(1+x \right) -\frac{x}{1+x}$, it is strictly increasing for $x>0$ by checking its first order derivative. Therefore, we have $\text{SNR}_n^T=\theta^{-1} (\frac{2\lambda}{\mu W})$, in which $\lambda$ and $\mu$ are  Lagrangian multipliers and remains the same for $\forall n\in \mathcal{N}$. Hence $\text{SNR}_n^T$ is a constant irrespective of $n$ and can be denoted as $\text{SNR}^T$.


\section{Proof of Lemma \ref{lem:tdma_equal}} \label{A:lem_tdma_equal}
For the holding of the equality in (\ref{e:tdma_lower_con_data}), we prove by contradiction.
Note that the right-hand side of (\ref{e:tdma_lower_con_data}) is monotonically increasing for both $E_n$ and $t_n$, $\forall n\in \mathcal{N}$, if $\{E_n^{\dagger}\}$ and $\{t_n^{\dagger}\}$ are optimal solution for the problem and lead to $d<\sum_{n=1}^N t_n^{\dagger} W\ln \left(1+\frac{E_n^{\dagger} h_n}{\sigma^2W t_n^{\dagger}} \right)$, one can always reduce objective value by randomly choosing $i\in \mathcal{N}$ and replacing $E_i^{\dagger}$ with $E_i^{\ddagger}$ such that $d=\sum_{n=1,n\neq i}^N t_n^{\dagger} W\ln \left(1+\frac{E_n^{\dagger} h_n}{\sigma^2W t_n^{\dagger}} \right)+t_i^{\dagger} W\ln \left(1+\frac{E_i^{\ddagger} h_i}{\sigma^2W t_i^{\dagger}} \right)$. This contradicts to the optimality of $\{E_n^{\dagger}\}$ and $\{t_n^{\dagger}\}$. Hence the equality in (\ref{e:tdma_lower_con_data}) should hold.

With the holding of the equality in (\ref{e:tdma_lower_con_data}), we assume the optimal solution $\{E_n^{\dagger}\}$ and $\{t_n^{\dagger}\}$ lead to $d=\sum_{n=1}^N t_n^{\dagger} W\ln \left(1+\frac{E_n^{\dagger} h_n}{\sigma^2W t_n^{\dagger}} \right)$ and $2\sum_{n=1}^N t_n^{\dagger} < T-\frac{Ld}{f_B}$. Here, we can randomly choose $i\in \mathcal{N}$ and replace $t_i^{\dagger}$ with $t_i^{\ddagger}$ such that $2\left(\sum_{n=1,n\neq i}^N t_n^{\dagger} +t_i^{\ddagger}\right)=T-\frac{Ld}{f_B}$. It is obvious that $t_i^{\ddagger}>t_i^{\dagger}$, which can lead to $E_i^{\ddagger}<E_i^{\dagger}$ due to the holding of (\ref{e:tdma_lower_con_data})'s equality. This can contribute to the decrease of Problem \ref{p:tdma_lower}'s cost function and contradicts to the optimality of $\{E_i^{\dagger}\}$ and $\{t_i^{\dagger}\}$. Therefore, the holding of the equality in (\ref{e:tdma_lower_con_time}) is proved.
	

\section{Proof of Lemma \ref{lem:tdma_upper_convex}} \label{A:lem_tdma_upper_convex}
To prove the convexity of Problem \ref{p:tdma_upper}, we need to prove that $U(d)$ is a convex function of $d$.
Define the right-hand side function of constraint (\ref{e:tdma_lower_final_con_data}) as
\begin{equation} \label{e:tdma_upper_data}
\sigma(d)= \frac{1}{2} \sigma^2W\left(T-\frac{Ld}{f_B} \right) \left(e^{\frac{2d}{W\left(T-\frac{Ld}{f_B} \right)}}-1 \right)
\end{equation}
whose second order derivative with $d$ is
\begin{equation} \label{e:tdma_upper_data_secdev}
\sigma''(d)= \frac{2\sigma^2f_B^3 T^2}{W\left(f_BT-Ld\right)^3} e^{\frac{2df_B}{W\left(f_BT-Ld\right)}}
\end{equation}
which is larger than zero since $f_B T\geq Ld$.
Therefore, $\sigma(d)$ is a convex function with $d$.
	
Suppose $\{E_n^{\dagger}\}$ and $\{E_n^{\ddagger}\}$ are the set of optimal solutions of $\{E_n\}$ for Problem \ref{p:tdma_lower_final} when $d=d^{\dagger}$ and $d=d^{\ddagger}$ respectively, i.e. $U(d^{\dagger})=\sum_{n=1}^N E_n^{\dagger}\left(1+\frac{h_n}{g_n} \right)$ and $U(d^{\ddagger})=\sum_{n=1}^N E_n^{\ddagger}\left(1+\frac{h_n}{g_n} \right)$. To satisfy constraint (\ref{e:tdma_lower_final_con_data}), there should be $\sum_{n=1}^N E_n^{\dagger}h_n\geq \sigma(d^{\dagger})$ and $\sum_{n=1}^N E_n^{\ddagger}h_n\geq \sigma(d^{\ddagger})$.
	
For $a \in [0,1]$, we have
\begin{equation}
	 a \sum_{n=1}^N E_n^{\dagger}h_n + (1-a) \sum_{n=1}^N E_n^{\ddagger}h_n
	\geq  a \sigma(d^{\dagger}) + (1-a) \sigma(d^\ddagger)
	\geq \sigma \left(a d^{\dagger}+(1-a) d^{\ddagger}\right)
\end{equation}
which means $E_n=a E_n^{\dagger} + (1-a) E_n^{\ddagger}$ for $n \in \mathcal{N}$ is a feasible solution for $d=a d^{\dagger}+(1-a) d^{\ddagger}$. In this case
	\begin{equation}
          U( a d^{\dagger} + (1-a) d^{\ddagger} )
	\leq   a \sum_{n=1}^N E_n^{\dagger} \left(1+\frac{h_n}{g_n} \right) + (1- a) \sum_{n=1}^N E_n^{\ddagger} \left(1+\frac{h_n}{g_n} \right)
	=  a U(d^{\dagger})+(1-a) U(d^\dagger).
	\end{equation}
Hence $U(d)$ is convex with $d$ and Problem \ref{p:tdma_upper} is a convex optimization problem.


\section{Proof of Lemma \ref{lem:fdma_equal}} \label{A:lem_fdma_equal}
Substitute $P_n=\frac{E_n}{t}$ into Problem \ref{p:fdma_deleteq} for $n\in \mathcal{N}$, (\ref{e:fdma_deleteq_con_data}) becomes $d\leq \sum_{n=1}^N tw_n\ln\left(1+\frac{E_nh_n}{t\sigma^2w_n} \right)$, and the cost function of Problem \ref{p:fdma_deleteq} turn to be $\sum_{n=1}^N E_n \left(1+\frac{h_n}{g_n} \right) +\frac{\kappa L^3 (D-d)^3}{T^2}$. We first prove the holding of (\ref{e:fdma_deleteq_con_data})'s equality via contradiction.
Suppose $d^{\dagger}$, $t^{\dagger}$, $\{E_n^{\dagger}\}$ and $\{w_n^{\dagger}\}$ are optimal solution and $d^{\dagger}<\sum_{n=1}^N t^{\dagger}w_n^{\dagger}\ln\left(1+\frac{E_n^{\dagger}h_n}{t^{\dagger}\sigma^2w_n^{\dagger}} \right)$, then we can randomly choose $i\in \mathcal{N}$ and replace $E_n^{\dagger}$ with $E_n^{\ddagger}$ that satisfies $d=\sum_{n=1,n\neq i}^N t^{\dagger}w_n^{\dagger}\ln\left(1+\frac{E_n^{\dagger}h_n}{t^{\dagger}\sigma^2w_n^{\dagger}} \right) +t^{\dagger}w_i^{\dagger}\ln\left(1+\frac{E_i^{\ddagger}h_i}{t^{\dagger}\sigma^2w_i^{\dagger}} \right)$, which is a feasible solution and can also result in less cost function of Problem \ref{p:fdma_deleteq}. This is contradictory to the assumption that $d^{\dagger}$, $t^{\dagger}$, $\{E_n^{\dagger}\}$ and $\{w_n^{\dagger}\}$ are optimal solution. Hence the holding of (\ref{e:fdma_deleteq_con_data})'s equality is proved.
	
Next we prove the holding of (\ref{e:fdma_deleteq_con_time})'s equality. With the equality of (\ref{e:fdma_deleteq_con_data}) active, there is $d=\sum_{n=1}^N tw_n\ln\left(1+\frac{E_nh_n}{t\sigma^2w_n} \right)$, whose right-hand side is a monotonically increasing function with respect to $t$, $E_n$, and $w_n$ for $n\in \mathcal{N}$. Therefore, to reduce the cost function of Problem \ref{p:fdma_deleteq}, we need to reduce $\{E_n\}$  as much as possible, which will lead to the increase of $t$ as much as possible, for given $d$ and $\{w_n\}$. When $t$ grows to its maximum, the equality of (\ref{e:fdma_deleteq_con_time}) is active.
Proof for the activeness of (\ref{e:fdma_deleteq_con_band})'s equality is similar with the one for the activeness of (\ref{e:fdma_deleteq_con_time})'s equality and is omitted here.
	

\section{Proof of Lemma \ref{lem:af_convergence}} \label{A:lem_af_convergence}
Define the objective function of Problem \ref{p:af_lower_geometric} as $y(q,s,\{\alpha_n\})$, and the objective function of Problem \ref{p:af_lower_iterative} as $\bar{y}^i(q,s,\{\alpha_n\})$.
To verify the convergence of SCA method,  the following inequality needs to be proved for a feasible point $(q^i, s^i, \{\alpha_n^i\})$ of Problem \ref{p:af_lower_geometric}
\begin{equation} \label{e:af_lowerbound}
\bar{y}^i(q^i,s^i,\{\alpha_n^i\}) \geq y(q^i,s^i,\{\alpha_n^i\})
\end{equation}
which indicates that the objective function of Problem \ref{p:af_lower_geometric} can serve as a lower bound for that of Problem \ref{p:af_lower_iterative} at the point $\left(q^i, s^i, \{a_n^i\}\right)$. In the following, the inequality in (\ref{e:af_lowerbound}) will be proved first.

The left-hand side of (\ref{e:af_lower_con_geometric_concave}) is joint convex with respect to the vector of $q$, $s$, and $\alpha_n$ for $n\in \mathcal{N}$.
According to the property of a convex function's first-order derivative in \cite{convex_book}, for any feasible point $\left(q^i, s^i, \{\alpha_n^i\} \right)$ of Problem \ref{p:af_lower_geometric}, there is
	\begin{equation}
	\begin{split}
	&\frac{2\sqrt{h_ng_n}e^{\alpha_n^i}}{\sum_{n=1}^N \sqrt{h_ng_n}e^{\alpha_n^i}} (\alpha_n -\alpha_n^i) +(q-q^i)-(s-s^i) +2\ln \left(\sum_{n=1}^N \sqrt{h_ng_n} e^{\alpha_n^i} \right)+q^i-s^i \\
	\leq & 2\ln \left(\sum_{n=1}^N \sqrt{h_ng_n} e^{\alpha_n} \right) +q-s.
	\end{split}\label{e:af_lower_con_iterative_lemma}
	\end{equation}
	Thus, (\ref{e:af_lower_con_iterative_concave}) is a sufficient but not necessary condition of (\ref{e:af_lower_con_geometric_concave}), and Problem \ref{p:af_lower_iterative} has smaller feasible region than Problem \ref{p:af_lower_geometric} at the point $\left(q^i, s^i, \{\alpha_n^i\} \right)$.
To this end,  Problem \ref{p:af_lower_geometric} may yield lower cost in terms of cost function, i.e., there is $\bar{y}^i(q^i,s^i,\{\alpha_n^i\}) \geq y(q^i,s^i,\{\alpha_n^i\})$, which is exactly the expression of (\ref{e:af_lowerbound}).

Since Problem \ref{p:af_lower_iterative} has smaller feasible region than Problem \ref{p:af_lower_geometric} at the point $\left(q^i, s^i, \{\alpha_n^i\} \right)$, which is feasible for Problem \ref{p:af_lower_geometric}, the associated optimal solution of Problem \ref{p:af_lower_iterative}, denoted as $\left(q^{i+1}, s^{i+1}, \{\alpha_n^{i+1}\} \right)$, is also feasible for Problem \ref{p:af_lower_geometric}.
Then there is
	\begin{equation} \label{e:af_convergence_better}
	y\left(q^i, s^i, \{\alpha_n^i\} \right)  = \bar{y}^i\left(q^i, s^i, \{\alpha_n^i\} \right)
	 \geq \bar{y}^i\left(q^{i+1}, s^{i+1}, \{\alpha_n^{i+1}\} \right)
	 \geq y\left(q^{i+1}, s^{i+1}, \{\alpha_n^{i+1}\} \right)
	\end{equation}
where the equality in the first line comes from the definition of Problem \ref{p:af_lower_iterative}, the inequality in the second line is due to the fact the $\left(q^{i+1}, s^{i+1}, \{\alpha_n^{i+1}\} \right)$ is the optimal solution of Problem \ref{p:af_lower_iterative} at point $\left(q^{i}, s^{i}, \{\alpha_n^{i}\} \right)$, and the inequality in the third line is from (\ref{e:af_lowerbound}).
	
The inequality in (\ref{e:af_convergence_better}) indicates that $\left(q^{i+1}, s^{i+1}, \{\alpha_n^{i+1}\} \right)$ yields cost in terms of cost function than $\left(q^{i}, s^{i}, \{\alpha_n^{i}\} \right)$ for Problem \ref{p:af_lower_geometric}. From Cauchy's theorem, in the sequence $\big \{\left(q^{i}, s^{i}, \{\alpha_n^{i}\} \right) \big \}$, there is a convergent subsequence $\big \{\left(q^{i_v}, s^{i_v}, \{\alpha_n^{i_v}\} \right)\big \}$ with limit point $\left(q^*, s^*, \{\alpha_n^*\} \right)$, such that
	$\lim_{v\to \infty} \left(y\left(q^{i_v}, s^{i_v}, \{\alpha_n^{i_v}\} \right) - y\left(q^*, s^*, \{\alpha_n^*\} \right)\right) =0.$
	For certain $i$, there must exists a $v$ satisfiying $i_v\leq i \leq i_{v+1}$, hence there is
	$y\left(q^{i_v}, s^{i_v}, \{\alpha_n^{i_v}\} \right)
	\geq y\left(q^{i}, s^{i}, \{\alpha_n^{i}\} \right)
	 \geq y\left(q^{i_{v+1}}, s^{i_{v+1}}, \{\alpha_n^{i_{v+1}}\} \right).$
	When $i$ goes to infinity, we have
	\begin{equation}
	\begin{split}
	0=&\lim_{v\to \infty} \left(y\left(q^{i_v}, s^{i_v}, \{\alpha_n^{i_v}\} \right) - y\left(q^*, s^*, \{\alpha_n^*\} \right)\right)
	\geq \lim_{i\to \infty} \left(y\left(q^{i}, s^{i}, \{\alpha_n^{i}\} \right) - y\left(q^*, s^*, \{\alpha_n^*\} \right)\right) \\
	\geq &\lim_{v\to \infty} \left(y\left(q^{i_{v+1}}, s^{i_{v+1}}, \{\alpha_n^{i_{v+1}}\} \right) - y\left(q^*, s^*, \{\alpha_n^*\} \right)\right) =0.
	\end{split}
	\end{equation}
Therefore, the sequence $\left(q^{i}, s^{i}, \{\alpha_n^{i}\} \right)$ is also convergent, with limit point being $\lim_{i\to \infty} \left(q^{i}, s^{i}, \{\alpha_n^{i}\} \right)=\left(q^*, s^*, \{\alpha_n^*\} \right)$. Based on Theorem 1 in \cite{SCA}, the limit point $\left(q^*, s^*, \{\alpha_n^*\} \right)$ would be a stationary point.
	
\vspace{-5mm}
\section{Proof of Lemma \ref{lem:af_upper_monotonic}} \label{A:lem_af_upper_monotonic}
The first-order derivative of $\psi(d)$ is
$\psi'(d)=\frac{2f_B^2T}{W\left(f_BT-Ld\right)^2} e^{\frac{2df_B} {W\left(f_BT-Ld\right)}}$
which is positive for $Ld\leq f_BT$. Then $\psi(d)$ is monotonic increasing. For Problem \ref{p:af_lower}, increasing $d$ will shrink the feasible region of Problem \ref{p:af_lower} and further increase the minimal cost function of Problem \ref{p:af_lower}, i.e., $X(d)$.


\end{document}